\documentclass[Royal,sagev,times]{sagej}

\usepackage{moreverb,url}
\usepackage{amsthm}
\theoremstyle{plain}
\newtheorem{theorem}{Theorem}
\usepackage{algorithm}
\usepackage{algpseudocode}
\usepackage{caption}
\usepackage{subcaption}

\usepackage[colorlinks,bookmarksopen,bookmarksnumbered,citecolor=red,urlcolor=red]{hyperref}

\newcommand\BibTeX{{\rmfamily B\kern-.05em \textsc{i\kern-.025em b}\kern-.08em
T\kern-.1667em\lower.7ex\hbox{E}\kern-.125emX}}

\DeclareMathOperator*{\argmax}{arg\,max}
\DeclareMathOperator*{\argmin}{arg\,min}

\def\volumeyear{2022}

\begin{document}

\runninghead{Watson, Girling, Hemming, and Lilford}

\title{Optimal Study Designs for Cluster Randomised Trials: An Overview of Methods and Results}

\author{Samuel I. Watson\affilnum{1}, Alan Girling\affilnum{1}, Karla Hemming\affilnum{1}}

\affiliation{\affilnum{1}University of Birmingham, Birmingham, UK} 

\corrauth{Samuel I. Watson, Institute for Applied Health Research,
University of Birmingham,
Birmingham,
B152TT, UK.}

\email{s.i.watson@bham.ac.uk}

\begin{abstract}
There are multiple possible cluster randomised trial designs that vary in when the clusters cross between control and intervention states, when observations are made within clusters, and how many observations are made at each time point. Identifying the most efficient study design is complex though, owing to the correlation between observations within clusters and over time. In this article, we present a review of statistical and computational methods for identifying optimal cluster randomised trial designs. We also adapt methods from the experimental design literature for experimental designs with correlated observations to the cluster trial context. We identify three broad classes of methods: using exact formulae for the treatment effect estimator variance for specific models to derive algorithms or weights for cluster sequences; generalised methods for estimating weights for experimental units; and, combinatorial optimisation algorithms to select an optimal subset of experimental units. We also discuss methods for rounding experimental weights, extensions to non-Gaussian models, and robust optimality. We present results from multiple cluster trial examples that compare the different methods, including determination of the optimal allocation of clusters across a set of cluster sequences and selecting the optimal number of single observations to make in each cluster-period for both Gaussian and non-Gaussian models, and including exchangeable and exponential decay covariance structures.
\end{abstract}

\keywords{Cluster randomised trial, optimal experimental design, generalised linear mixed model}

\maketitle

\section{Introduction}
The cluster randomised trial is an increasingly popular experimental study design. It is used to evaluate interventions applied to groups of people, like classrooms, clinics, or villages, or when the outcome for one individual in the group depends on the outcomes for the others, as is the case with infectious diseases, for example \cite{Eldridge2012,Murray1998}. The design of a cluster trial involves the specification of all aspects of the study, many of which are determined by practical, ethical, and contextual restrictions. However, one major aspect of cluster trial design that can be resolved, or at least supported, with statistical analysis is the sample size of individuals and clusters, when observations are captured from the clusters and individuals, and when each cluster receives the intervention(s). 

From both an ethical and practical standpoint, minimising the number of individuals, clusters, or observations required to achieve an inferential goal is highly desirable. For cluster trials, inferences are almost always based on the variance of the estimator of the treatment effect. Designs that minimise the variance of a specific parameter, or combination of parameters, are said to be `\textit{c}-optimal'. However, for any particular design problem, enumerating all the different possible designs and their associated variances to identify the \textit{c}-optimal design is often impossible, given the number of possible variants. Therefore, we use algorithms that can quickly identify an efficient, or `optimal', design. The correlation of outcomes within clusters, over time, and potentially within-individuals over time, makes the analysis of the efficiency of a study design more complex though, over and above individual-level studies with independent observations.\cite{Hooper2016,Hemming2020,Li2021} 

There has been recent methodological advances in the optimal experimental design literature to identify optimal designs in studies with correlated observations, as well as several recent studies to look at the problem specifically for certain types of cluster randomised trial. In this article, we review the literature on optimal cluster randomised trial designs and review and translate more general methods and algorithms from the broader literature to this context. We present results for different cluster trial design scenarios using a range of methods to illustrate use of the different approaches and to identify optimal cluster trial design for a range of contexts. 

\section{What is the optimal cluster trial problem?}
There are multiple types of optimality in the experimental design literature, which are referenced using a ``alphabet'' system of letters.\cite{Atkinson2007,Berger2009} The primary objective of a cluster randomised trial is almost always to provide an estimate of the treatment effect of an intervention and an associated measure of uncertainty for one or more outcomes. Other parameters in the statistical model, such as the covariance parameters or intraclass correlation coefficient, are not of primary interest. For example, the predominant method used to justify the sample size within a particular study design is the power for a null hypothesis significance test of the treatment effect parameter \cite{Hooper2016,Hemming2020}. Thus, efficiency and optimality in this setting relates to minimising the variance of the treatment effect estimator, which is \textit{c}-optimality.

We now introduce concepts and notation to describe the methods to identify \textit{c}-optimal cluster trial designs. We assume time is modelled discretely where there are repeated measures, such that observations are considered to have been made within clusters at discrete points in time. Approximations can be made to a continuous time model by finely discretising time within this framework; using a regular grid over a continuous space is a common strategy in optimal design work.\cite{Yang2013} In what follows, we represent matrices using capital letters, e.g. $X$. We use a subscript to denote submatrices or elements of a matrix, in particular, we notate $X_A$ as the rows of $X$ in set $A$ with all the columns, and $\Sigma_{A}$ as the sub-matrix of $\Sigma$ with rows and columns in $A$. Lower case letters represent scalars and so $X_{i,j}$ indicates the element of $X$ in the $i$th row and $j$th column.

We assume there are $N$ possible observations that could be made as part of the study. An observation consists of a single `design point' that generates an outcome datum. Often observations are grouped into higher-level units around which the study is designed. For cluster trials with cross-sectional sampling each observation will be a unique individual grouped into cluster-periods and then clusters. For cohort designs, observations will be grouped within an individual trial participant and then in clusters. We refer to an \textit{experimental unit} as the smallest indivisible set of observations for the design problem. For example, we may wish to choose which whole cluster sequences to include, so the cluster sequence is the experimental unit. Equally, we may select cluster-periods, if we do not need to include all time periods within a cluster, or indeed which specific observations. The set of all experimental units is the \textit{design space}. To simplify matters, we assume each observation is in one and only one experimental unit. However, experimental units in the design space need not be unique. For example, in the absence of individual-level covariates, an observation made in a cluster at a given time will have identical values of the fixed effect parameters to all other observations in the same cluster-period.

\begin{figure}
    \centering
    \includegraphics[width=\textwidth]{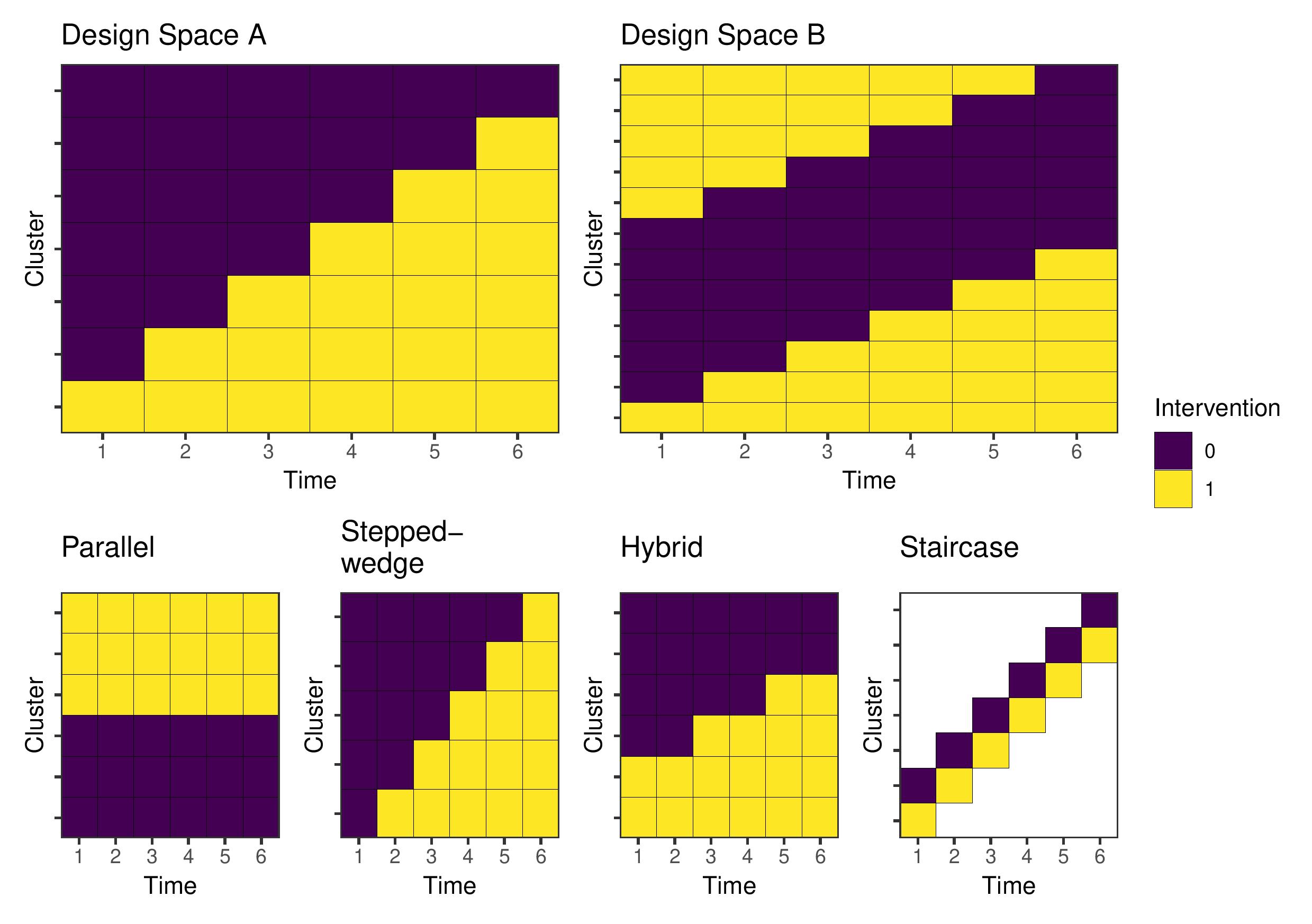}
    \caption{Examples of cluster trial design spaces and study designs for six time periods. Each row represents a cluster, or cluster sequence, and may be repeated more than once in the design space. Each cell is a cluster-period and contains one or more individual potential observations. Design Space A encodes a no reversibility assumption and includes contemporaneous comparisons. Design Space B allows for both addition and removal of the intervention over time. Parallel, stepped-wedge, hybrid, and staircase are all designs within both design spaces.}
    \label{fig:desspace}
\end{figure}

The top left panel of Figure \ref{fig:desspace} (Design Space A) represents a cluster randomised trial design space. This design space includes the following restrictions:
\begin{enumerate}
    \item[i] No reversibility, i.e. clusters can only cross from control to intervention states. 
    \item[ii] There must be contemporaneous comparison in at least one time period, i.e. a before-and-after design would not be permitted since it would not include a randomised comparison.
\end{enumerate}
Each row indicates a cluster sequence, and each column a discrete time interval or period. Within each cell, there are a pre-specified number of observations. Each row in the diagram is illustrated only once, but there may be multiple repeats of the same row in the design space depending on the formulation of the problem. Where there are multiple instances of the same cluster sequence, the design space includes the most common types of cluster randomised trial design with repeated measures: a parallel design, in which a cluster receives the intervention in all periods or the control in all periods, or a stepped-wedge cluster randomised trial, where the intervention roll out is staggered such that all clusters start in the control condition and then one or more clusters receives the intervention in each time period until all clusters are in the intervention state. A `hybrid' design consists of a mix of parallel and stepped-wedge cluster sequences, and a `staircase' design includes only the cluster-periods on the diagonal. Figure 1 illustrates these designs. Given that this design space incorporates among the most widely used cluster randomised trial designs, and that these restrictions reflect common real-world limitations, it is an obvious choice for many applications. However, more complex design spaces (such as Design Space B) are required to allow for alternative designs like cluster cross-over. Such a design space is illustrated in Figure 2, which removes the no reversibility restriction. In these cases the cross-over design is almost always the optimal design.\cite{Girling2016}

We base our analyses around a generalised linear mixed model (GLMM). For outcome vector $y$:
\begin{align}
    \begin{split}
    \label{eq:glmm}
        y &\sim F(\mu,\sigma) \\
        \mu &= h^{-1}(\eta)\\
        \eta &= X\beta + Z\mathbf{u}\\
        \mathbf{u} &\sim N(0,D)
    \end{split}
\end{align}
where $F$ is a statistical distribution with mean $\mu$ and scale parameter $\sigma$, and $h^{-1}$ is a link function. We assume that the distribution $F$ is in the exponential family. The matrix $X$ is the design matrix of fixed effects, the matrix $Z$ is the design matrix for the random effects $\mathbf{u}$, and $D$ is the covariance matrix of the random effects. We discuss below the specification of $X$, $Z$, and $D$.

We assume that the matrices $X$ and $Z$ have $N$ rows and so contain all of the observations in the design space. There are $J$ experimental units, which we denote as $\mathcal{E}_j$ for $j=1,...,J$ where each $\mathcal{E}_j \subset [1,...,N]$ contains a subset of the rows. We also denote the design space as $\mathcal{D} := \{\mathcal{E}_j : j=1,...,J \}$ and a specific design as $d \subset \mathcal{D}$. Our aim is to identify the `optimal' design $d^*$ of size $m<J$ by selecting the most efficient set of $m$ experimental units from $\mathcal{D}$.

Most experimental design criteria are based on the Fisher information matrix. For the GLMM above, the information matrix for the generalised least squares estimator, the best linear unbiased estimator, for a particular design is:
\begin{equation}
\label{eq:infomat}
    M_d = X_d^T \Sigma_{d}^{-1} X_d
\end{equation}
where $\Sigma$ is the covariance matrix of the observations $y$. Then the \textit{c}-optimal design criterion is:
\begin{equation}
\label{eq:coptim}
    f(d) = \begin{cases}
        c^T M_d^{-1} c & \text{if } M \text{positive semi-definite} \\
        \infty & \text{ otherwise}
    \end{cases}
\end{equation}
where $c$ is a vector consisting of zeroes except for a one in the position of the treatment effect parameter. For some designs, such as if there were no observations in the treatment condition, $M$ would not be positive semi-definite and so we define the variance as infinite, i.e. the design provides no information on the parameter. The formal design problem is then to find $d^* \subset \mathcal{D}$ that minimises $f$ such that $\vert d^* \vert = m < J$, i.e. $d^* = \argmin_d f(d)$.

\subsection{GLMM Specifications for Cluster Randomised Trials}
Without loss of generality, we focus on models for cluster trials where individuals are cross-sectionally sampled in each cluster-period. Where relevant and also without loss of generality, we use $r$ to represent the number of observations per cluster-period. For a comprehensive discussion of different models relevant to cluster randomised trials, see Li et al\cite{Li2021}. 

\subsubsection{Covariance Function}
The observed outcome for an individual $i$ in cluster $k$ at time $t$ is specified as $y_{ikt}$ with linear predictor $\eta_{ikt} = x_{ikt}\beta + s_{ikt}$ where $s_{ikt} = z_{ikt}\mathbf{u}$ represents the random effects. The covariance function defines the entries of the covariance matrix $D$. We define a covariance function as:
\begin{equation}
\label{eq:covfunc}
    \text{Cov}(s_{ikt},s_{i'k't'}) = g(\Delta t, \Delta k)
\end{equation}
where $\Delta t = \vert t - t' \vert$ and $\Delta k = \vert k - k' \vert$. We define the following covariance functions:
\begin{enumerate}
    \item[EXC1] Cluster Exchangeable \begin{equation*}
        g(\Delta t, \Delta k) = \begin{cases}
            \tau^2 & \text{ if } \Delta k = 0 \\
            0 & \text{ otherwise}
        \end{cases}
    \end{equation*}
    \item[EXC2] Nested Exchangable \begin{equation*}
        g(\Delta t, \Delta k) = \begin{cases}
            \tau^2 + \omega^2 & \text{ if } \Delta k = 0 \text{ and } \Delta t = 0 \\
            \tau^2 & \text{ if } \Delta k = 0 \text{ and } \Delta t > 0 \\
            0 & \text{ otherwise}
        \end{cases}
    \end{equation*}
    \item[AR1] Auto-regressive or exponential decay \begin{equation*}
        h(\Delta t, \Delta k) = \begin{cases}
            \tau^2 \lambda^{\Delta t} & \text{ if } \Delta k = 0 \\
            0 & \text{ otherwise}
        \end{cases}
    \end{equation*}
\end{enumerate}
In the cluster and nested exchangeable functions above, the parameter $\tau^2$ represents the between cluster variance, $\omega^2$ is the within-cluster, between-period variance. For the auto-regressive function, $\tau^2$ similarly represents the between cluster variance, with $\lambda$ the auto-regressive parameter describing the rate of temporal decay.

For Gaussian-identity models, we use $\sigma^2$ to denote the observation-level variance. Gaussian-identity models are often re-parameterised in terms of other parameters, in particular:
\begin{itemize}
    \item[ICC] Intra-class correlation coefficient. Equal to $\rho = \frac{\tau^2}{\tau^2 + \sigma^2}$ for EXC1 and AR1 and $\rho = \frac{\tau^2 + \omega^2}{\tau^2 + \sigma^2 + \omega^2}$ for EXC2. More precisely, this is the `within-period ICC' for designs with repeated measures and EXC2 function.\cite{Hooper2016}
    \item[CAC] Cluster-autocorrelation coefficient. Equal to $r = \frac{\tau^2}{\tau^2 + \omega^2}$ for EXC2 and not defined for the other models. 
\end{itemize}

\subsubsection{Matrix X}
The $n \times P$ matrix $X$ is a matrix of covariates. For cluster trials with repeated measures $X$ typically consists of an intercept, time period indicators for $T-1$ time periods, and a treatment indicator. Equivalently, matrix $X$ may be specified without the intercept and with $T$ time period indicators. We use this specification for the examples in subsequent sections. For some trial designs, investigators may consider alternative specifications. For a parallel design, the treatment effect estimator from a model that does not adjust for time period is unbiased. However, such a specification would result in a biased treatment effect estimator where the intervention roll out is staggered over time. Thus, we assume that for most applications where staggered designs feature in the design space, adjustment for time is incorporated in the specification for $X$. We may also consider adjusting for continuous functions of time, such as polynomials, if the time periods are an approximation to continuous time. For example, Hooper and Copas\cite{Hooper2021} consider cubic and piecewise continuous polynomials.

In this discussion, we also assume there is a single treatment that enters the model as a dichotomous treatment indicator. More complex cluster trial designs may feature multiple arms and treatments,\cite{Watson2021} including continuous treatments representing dose. We do not consider these designs here, however the optimal design methods below can be extended to these cases.

\section{Methods and previous literature}
We divide the currently available methods for the optimal cluster trial design problem into three categories: (i) derivation of exact formulae for the treatment effect variance or precision for specific models and design spaces; (ii) general `multiplicative' methods that derive weights to place on each unique experimental unit; and (iii) general combinatorial optimisation algorithms designed to select the optimum $m$ items from a discrete set of size $J$.

\subsection{Exact Formulae}
For simpler models one can derive explicit formulae for $f(d)$. Given a statement of the variance or precision one can then either determine an algorithm to identify an optimal solution, or use it to calculate the variance for a wide range of designs and/or parameter values and compare numerically or graphically. Many such studies are based on the formula for the variance of the treatment effect estimator in the linear mixed model with EXC1 covariance given by Hussey and Hughes.\cite{Hussey2007}

Girling and Hemming\cite{Girling2016} provide perhaps the most notable study of this type for cluster trials. They derive a formula for the precision of the treatment effect estimator in a linear mixed model under covariance structure EXC2 along with individual-level cohort effects, although we drop the individual level cohort effects for this summary. They consider the problem of determining which periods to introduce the intervention into each of the $m$ clusters. Each cluster is observed in each of the $T$ time periods, and each cluster-period has $n$ observations. One can map this problem onto Design Space A in Figure \ref{fig:desspace} where each row is an experimental unit repeated $m$ times, and the goal is to identify the optimal $m$ set of experimental units.

We can rewrite model (\ref{eq:glmm}) as a linear mixed model for individual $i$ in cluster $k$ at time $t$ as:
\begin{equation}
\label{eq:lmm2}
    y_{ikt} = J_{kt}\delta + w_t\gamma_t + \alpha_{k} + \theta_{kt} + u_{ikt}
\end{equation}
where $J_{kt}$ is an indicator for if cluster $k$ has the intervention at time $t$, $w_t$ is a time period indicator with time period parameters $\gamma_t$, and $\alpha_{1k} \sim N(0,\tau^2)$ and $\theta_{kt} \sim N(0,\omega^2)$ are the cluster and cluster-period random effect terms, and $u_{ikt} \sim N(0,\sigma^2)$ is the error term. The fixed effect parameters are $\beta = [\delta, \gamma_1, ..., \gamma_T]^T$ and $\delta$ is the treatment effect parameter. With discrete clusters and time periods, we can aggregate model (\ref{eq:lmm2}) into a model for the cluster-period means:
\begin{equation}
\label{eq:lmm3}
    \Bar{y}_{kt} = J_{kt}\delta + w_t\gamma_t + \alpha_{k} + e_{kt}
\end{equation}
where $\Bar{y}_{kt} = \frac{1}{r}\sum_{i=1}^r y_{ikt}$ is the mean outcome for cluster $k$ in time period $t$ and $\text{Var}(e_{kt}) \sim N(0, \omega^2 + \frac{\sigma^2}{r})$. Girling and Hemming, following work by Hussey and Hughes\cite{Hussey2007} and others, then show that the precision of the treatment effect estimator $\Hat{\delta}$ is given by:
\begin{align}
\begin{split}
\label{eq:prec}
    \frac{1}{f(d)} &= \text{Var}^{-1}(\Hat{\delta}) \\
    &= \frac{mT}{(\omega^2 + \frac{\sigma^2}{n})(1-\Bar{\rho})}(a_d - b_d R)
\end{split}
\end{align}
where $\Bar{\rho} = \frac{\tau^2}{\tau^2+\omega^2+\sigma^2/n}$ is equivalent to the ICC at the cluster-period mean level and $R = \frac{T\Bar{\rho}}{1+(T-1)\Bar{\rho}}$ is the cluster mean correlation. The coefficients $a_d$ and $b_d$ are determined by the study design:
\begin{align*}
    a_d &= \frac{1}{mT} \sum_{t=1}^T \sum_{k=1}^m (J_{kt} - \Bar{J}_{\cdot t})^2 \\
    b_d &= \frac{1}{m} \sum_{k=1}^m (\Bar{J}_{\cdot t} - \Bar{J}_{\cdot \cdot})^2
\end{align*}
where the dot indicates the index over which the mean is taken. 

Girling and Hemming\cite{Girling2016} provide a method for using Equation (\ref{eq:prec}) to produce an optimal design under a no reversibility constraint. We assume the clusters are numbered such that a lower numbered cluster has greater than or equal number of intervention periods than any higher numbered cluster. We then map the cluster-period indexed to coordinates on a unit square $(j,t) \mapsto (x_{0j},x_{1t})$ where $x_{0j}, x_{1t} \in [-1/2,1/2]$. All cluster-periods start in the control state, and then starting with cluster 1 in the $T$th period, one successively changes the cluster-period to an intervention state in the order of decreasing values of $R x_{1t} - x_{0j}$ until $\Bar{J}_{\cdot \cdot}$ cluster-periods are included in the treated set. Examples of this method are provided in the article, which we reproduce in the examples section.

Lawrie, Carlin, and Forbes\cite{Lawrie2015} derive explicit formulae for the optimal proportion of clusters to allocate to each sequence (row) in the Design Space A to minimise $f(d)$ using a linear mixed model and the EXC1 covariance function. They show that the optimal proportion of clusters allocated to the $t$th sequence in the stepped-wedge design space (see Figure \ref{fig:desspace}) with $T-1$ sequences, $\phi_{t}$ is:
\begin{align}
\begin{split}
\label{eq:probweight1}
    \phi_1 &= \phi_{T-1} = \frac{1 + \rho(3r-1)}{2(1+\rho(rT-1)} \\
    \phi_t &= \frac{r\rho}{1+\rho(rT-1)} \text{ for } t=2,...,T-2
    \end{split}
\end{align}
A similar analysis for EXC1 structure and a linear mixed model is given in Woertman et al.\cite{Woertman2013}

Zhan, Bock, and Heuvel\cite{Zhan2018} extend Lawrie et al's analyses using Girling and Hemming's work to identify more general `optimal unidirectional switch designs' by extending the probability weights (\ref{eq:probweight1}) to a larger design space with sequences incorporating exclusively control or intervention conditions, and with EXC1 covariance functions. Here, unidirectional switching means  no reversibility, giving, for example Design Space A in Figure \ref{fig:desspace}. The more general probability weights for the design space with $T+1$ sequences are:
\begin{align}
\begin{split}
\label{eq:probweight2}
    \phi_0 &= \phi^{T} = \frac{1 + \rho(r-1)}{2(1+\rho(rT-1)} \\
    \phi_t &= \frac{r\rho}{1+\rho(rT-1)} \text{ for } t=1,...,T-1
    \end{split}
\end{align}
Zhan, Bock, and Heuvel also extend this analysis to smaller design spaces including only a subset of the rows of Design Space A. We discuss below methods for rounding proportions to whole numbers of clusters. 

There are several other studies that derive expressions for the treatment effect variance to identify efficient study designs. Hooper and Copas\cite{Hooper2021b} consider a linear mixed model with AR1 covariance for a cluster randomised trial with continuous recruitment. They consider a parallel study design with baseline measures and aim to identify the when the intervention should be implemented in the intervention arm under different sample sizes and covariance parameters. They calculate the value of (\ref{eq:coptim}) for a large range of models and graphically compare the results. Copas and Hooper\cite{Copas2020} take a similar approach with a linear mixed model with EXC1 covariance with a parallel trial design. They aim to identify optimal sample sizes and the proportion of data to collect in baseline and endline periods. Moerbeek\cite{Moerbeek2020} also uses an explict criterion, although not strictly for c-optimality, as they aim to identify an optimal sample size within treatment and control groups subject to a budget constraint. They consider only a single time period, such that the treatment effect estimator is a difference in means. Lemme et al also consider a similar cost-benefit optimisation approach for multicentre trials.\cite{Lemme2018}

Deriving explicit formulae for the variance or precision is appealing due to its relative simplicity. Identifying a c-optimal design does not require specialist tools and can be done using spreadsheet software. However, these methods are typically limited to specific models and designs, such as exchangeable covariance structures, linear models, and equal cluster-period sizes. The mathematical approach used to derive the precision formula does not carry over to more complex covariance structures or design spaces, nor to problems where the experimental unit is an observation or cluster-period. One can calculate the value of the c-optimality criterion directly for any design, as Hooper and Copas\cite{Hooper2021b} do. However, the number of designs one must calculate the variance for grows exponentially and prohibitively with the size of the design space. More general methods are required for these extended problems.

\subsection{Multiplicative weighting methods}
Determining probability weights for experimental units, as the studies cited above do explicitly\cite{Zhan2018,Lawrie2015}, is a useful strategy to simplify the optimal design problem. One can generalise this approach to tackle more complex models and design spaces. We place a probability measure $\phi$ on $\mathcal{D}$ so that our design is characterised by $\boldsymbol{\phi} := \{(\mathcal{E}_j,\phi_j):j=1,...,J\}$ where $\phi_j \in [0,1]$ are weights. The optimal design problem can then be re-stated as finding a design that minimises $f(\boldsymbol{\phi})$.

\subsubsection{Elfving's Theorem}
Elfving's Theorem is a classic result in the theory of optimal designs.\cite{Elfving1952} The original formulation considered independent, identically distributed observations. Holland-Letz, Dette, and Pepelyshev (2011)\cite{Holland-Letz2011} and Sangol (2011)\cite{Sagnol2011} generalised the theorem to the case where there is correlation within experimental units and multiple observations, such as within a cluster, but not \textit{between} experimental units, such as if the experimental unit was a cluster-period or observation. Elfving's theorem provides a geometric characterisation of the \textit{c}-optimal design problem. If the experimental units are uncorrelated, the information matrix in Equation (\ref{eq:infomat}) can be rewritten as:
\begin{equation*}
    M_d = \sum_{\mathcal{E}_j \in d} X^T_{\mathcal{E}_j}\Sigma^{-1}_{\mathcal{E}_j}X_{\mathcal{E}_j}
\end{equation*}
thus for the approximate design $\boldsymbol{\phi}$ we can write:
\begin{equation}
\label{eq:infomatsum2}
    M_{\boldsymbol{\phi}} = \sum_{k=1}^K X_{\mathcal{E}_k} \Sigma_{\mathcal{E}_k}^{-1} X_{\mathcal{E}_k} \phi_k
\end{equation}
 which we can rewrite as:
\begin{equation}
\label{eq:infoF}
      M_{\boldsymbol{\phi}} = \sum_{k=1}^K F_{\mathcal{E}_k}^T F_{\mathcal{E}_k} \phi_k
\end{equation}
where $F_{\mathcal{E}_k} = L_{\mathcal{E}_k,\mathcal{E}_k}^T X_{\mathcal{E}_k}$ and $L_{\mathcal{E}_k,\mathcal{E}_k}$ is a square root of $ \Sigma_{\mathcal{E}_k,\mathcal{E}_k}^{-1}$.

A `generalised Elfving set' is:
\begin{equation}
    \mathcal{R} = \text{co}\{   F_{\mathcal{E}_{k}}^T \epsilon_k: X_{\mathcal{E}_k} \in \mathcal{X}^{\vert \mathcal{E}_k \vert \times P}; || \epsilon_k ||= 1; k = 1,...,K \}
\end{equation}
where $\text{co}$ denotes the convex hull. This set leads us to a generalised Elfving theorem:
\begin{theorem}[Generalised Elfving Theorem]\label{th2}
A design $\boldsymbol{\phi} := \{(\mathcal{E}_k,\phi_k):k=1,...,K\}$ is c-optimal if and only if there exists vectors $\epsilon_1,...,\epsilon_K$ where $||\epsilon_k|| = 1$ and positive real scalar $\pi$ such that $\pi c = \sum_{k=1}^K \phi_k F_{\mathcal{E}_k}^T \epsilon_k$ is a boundary point of the set $\mathcal{R}$.
\end{theorem}
For proof see\cite{Holland-Letz2011,Sagnol2011}.

Sagnol\cite{Sagnol2011} shows how the generalised Elfving theorem can be used to define a second-order cone program, which is a type of conic optimisation problem than can be solved with interior point methods. This program returns the optimal values of $\phi_1,...,\phi_k$. We provide functionality for the problems we consider in this article in the R package \texttt{glmmrOptim}, including this program. Other proposals exist for identifying the optimal weights, such as using a multiplicative algorithm based on an upper bound for the solution.\cite{HollandLetz2012}.

\subsubsection{Mixed Model Weights}
Girling (forthcoming) has proposed an algorithm for finding the optimum set of weights that can be applied to the case when experimental units are equivalent to cluster-periods. Since observations in this context are exchangeable within a cluster-period, when the weights are rounded to number of observations (see next section), the result is equivalent to when the experimental unit is a single observation. We consider the aggregated cluster-period model (\ref{eq:lmm3}). The best linear unbiased estimator for the linear combination $b = c^T\beta$ can be written as
\begin{align*}
    \hat{b} &= \mathbf{a}^T\mathbf{y} \\
    &= \mathbf{a'}^TL\mathbf{y} 
\end{align*}
where $\mathbf{a} = [a_{1,1},...,a_{1,T},a_{2,1},...,a_{K,T}]$ is a vector of weights, with $a_{k,t}$ the estimation weight for cluster $k$ and time $t$, and $\mathbf{a} = L^T\mathbf{a'}$. As before, $L$ is the Cholesky decomposition of $\Sigma$ and $F = L^TX$. By the Gauss-Markov theorem, the estimator is unbiased if $F^Ta = c$ for $a = F(F^TF)^{-1}c$. So we have that $a = \Sigma^{-1}X(X^T \Sigma^{-1}X)^{-1}c$, giving us the generalised least squares estimator. In the linear model case we can write $\Sigma = \frac{\sigma^2}{N}W + ZDZ^T$, where $W$ is a diagonal matrix of weights such that the number of observations in the cluster period $(k,t)$ is $\phi_{k,t}N$ with $\sum_{kt}\phi_{kt} = 1$. The variance of the estimator can then be written as:
\begin{align*}
    \text{Var}(\hat{b}) &= \mathbf{a}^T\Sigma \mathbf{a} \\
    &= \sum_{k=1}^K \mathbf{a}_k^T\Sigma_k \mathbf{a}_k \\
    &= \sum_{k=1}^K \sum_{t=1}^T \sum_{s=1}^T a_{kt}a_{ks}g(\vert t - s \vert, 0) + \frac{\sigma^2}{N}\sum_{k=1}^K \sum_{t=1}^T \frac{a_{kt}^2}{\phi_{kt}}
\end{align*}
where $g(.)$ is the covariance function (\ref{eq:covfunc}). Ignoring the first part of the final line, which is not determined by the cluster-period weights in each cell, the Cauchy-Schwarz inequality shows that:
\begin{equation*}
    \frac{\sigma^2}{N}\sum_{k=1}^K \sum_{t=1}^T \frac{a_{kt}^2}{\phi_{kt}} \geq \frac{\sigma^2}{N}\left(\sum_{k=1}^K \sum_{t=1}^T \vert a_{kt} \vert \right)^2
\end{equation*}
which then gives us a lower bound on the variance when adding in the coviarance terms. This inequality becomes an equality, and hence the minimal variance, when the weights are set as:
\begin{equation*}
    \phi_{kt} = \frac{\vert a_{kt} \vert}{\sum_{kt} \vert a_{kt} \vert}
\end{equation*}
The above argument therefore suggests a simple algorithm to identify the cluster-period weights than minimise the variance, and hence are the c-optimal design, which is shown in Algorithm \ref{alg:girling}. We have implemented this algorithm in the R package \texttt{glmmrOptim} as the ``Girling algorithm''. To ensure the algorithm terminates, we had to add additional steps to the algorithm described below. In particular, on each iteration of the algorithm we exclude cluster-periods where the weight is smaller than some lower bound ($10^{-7}$) to avoid the weights continually shrinking and causing possible floating point errors; and excluding time periods in the linear predictor if the total weights for that period are zero. The algorithm applies to only the cases where the experimental units are cluster-periods or individual observations; however, one might sum the weights within larger experimental conditions for different contexts.

\begin{algorithm}
\caption{Optimal mixed model weights for $J$ experimental units with a target total number of observations $N$ and $\epsilon$ is the tolerance of the algorithm.}
\label{alg:girling}
\begin{algorithmic}
\Procedure{Optimal mixed model weights}{}
 \State Let $\boldsymbol{\phi} = [\phi_1,...,\phi_J]$ with $\phi_j = 1/J$ for all $j$
 \State Set $\delta = 1$
 \While{$\delta > \epsilon$}
 \State $a \leftarrow \Sigma^{-1}X(X^T \Sigma^{-1}X)^{-1}c$
 \ForAll{$j \in \{1,...,J\}$}
    $\phi'_{j} \leftarrow \frac{\vert a_{j} \vert}{\sum_{j} \vert a_{j} \vert}$
 \EndFor
 \State $\delta \leftarrow \argmax_j \vert \phi_j - \phi'_j \vert$
 \ForAll{$j \in \{1,...,J\}$}
    $\phi_{j} \leftarrow \phi'_j$
 \EndFor
 \State $\Sigma \leftarrow (\sigma^2/N)\text{diag}(\boldsymbol{\phi}^{-1}) + ZDZ^T$ 
 \EndWhile
\EndProcedure
\end{algorithmic}
 \end{algorithm}

\subsection{Rounding proportions of experimental units}
Where a method produces an optimal design in terms of the proportion of experimental units of each type to include, we must use a rounding procedure to translate it into exact numbers. There are several methods for converting proportions to integer counts that sum to a given total. The problem was famously identified for converting popular vote totals in states into numbers of seats in the US House of Representatives; the solutions are named after their proposers.\cite{Balinski2002}  Pukelsheim and Rieder (1992)\cite{PUKELSHEIM1992} following others\cite{Fedorov1972} argue that the procedure of John Quincy Adams is the most efficient method of rounding to an exact design. As Pukelsheim and Rieder note though, the design weights do not contain enough information to exactly identify a experimental design, and so multiple designs may be generated. However, this procedure is based on the assumption that a `fair' allocation includes at least one experimental unit of each type. For many cluster trial design problems we do not require this restriction, for example, a parallel trial is optimal in some cases.\cite{Girling2016} In other cases though, there may be practical reasons to ensure staggering of the roll-out,\cite{Hooper2021,Hemming2015} in which case this rounding scheme would be the most efficient. Hamilton's rounding procedure is an alternative method. We initially assign $\lfloor J\phi_j \rfloor$ clusters to each sequence (where $\lfloor x \rfloor$ is the floor of $x$), and the incrementally add clusters according to the largest remainder $J\phi_j - \lfloor J\phi_j \rfloor$. In the later examples (and the implementation in the R package \texttt{glmmrOptim}) we use all rounding procedures and then select the design with the smallest value of $f(d)$ since evaluating the variance for the small number of possibly optimal designs does not bear a high computational cost.

While the solutions generated by different rounding schemes, and the algorithms discussed in the next section, may in fact be an exact optimal solution, they cannot guarantee such a result. In the results section we provide several examples where the results of the methods may disagree. The equivalence theorem\cite{Pukelsheim2006} provides precise conditions to check whether a given design is indeed optimal. However, it requires knowledge of the optimal design. Girling and Hemming\cite{Girling2016} use an approach of comparing the relative efficiency of the design to that of a cluster cross-over, which is the most efficient if it is within the design space. Not all design spaces include the cluster cross-over design, and so the optimal design may not be known. Holland-Letz, Dette, and Renard (2012)\cite{HollandLetz2012} derive a lower bound for the relative efficiency of a given design in the context of a pharmacokinetic study with correlated observations.

\subsection{Combinatorial Optimisation Algorithms}
Watson and Pan\cite{Watson2022} show how the \textit{c}-optimal design criterion in Equation (\ref{eq:coptim}) is a `monotone supermodular function', which means it is amenable to one of several combinatorial optimisation algorithms that are well-studied in the literature. A supermodular function is one for which, given a design $d \subset \mathcal{D}$ and a smaller design $d' \subseteq d$, then $f(d \cup \mathcal{E}) - f(d) \geq f(d' \cup \mathcal{E}) - f(d')$ is true.\cite{Sviridenko2017} Intuitively, one can see this is the case for the design problems considered in this article since it states that the decline in variance from adding a new experimental unit $\mathcal{E}$ is smaller for larger designs. The function is monotone decreasing if $d' \subseteq d \rightarrow f(d') \geq f(d)$, which means that the variance will be at least as large if you remove any observations. The advantage of these algorithms is that they allow identification of optimal designs in cases where there is correlation between experimental units, such as when the experimental units are cluster-periods or single observations.

The three algorithms relevant to supermodular function minimisation are the local search, the greedy search, and the reverse greedy search algorithm.\cite{Wynn1970,Fedorov1972,Fisher1978,Nemhauser1978} We exclude the greedy search algorithm here, as it starts from the empty set and successively adds observations. As we require a minimum of $P$ observations to ensure a positive semidefinite information matrix, the algorithm therefore performs poorly as Watson and Pan\cite{Watson2022} show. The local and reverse greedy searchers are shown in Algorithm box \ref{alg:combin}. These algorithms are also implemented in the R package \texttt{glmmrOptim}. 

Finding the subset of size $m$ from the design space that minimises $f(d)$ is an NP-hard problem, however, much work has been produced from the 1970s onwards on computationally efficient methods of finding approximate solutions. In some cases, these algorithms give a `constant factor approximation', that is the worst case result has a provable bound on $f(d)/f(d^*)$ if $d^*$ is the \textit{c}-optimal design. For the design problem we consider in this article, only the local search has a constant factor approximation. However, we also include the reverse greedy search as it, or similar variants, have appeared in the literature for cluster trials.

\begin{algorithm}
\caption{Combinatorial algorithms to generate a design of size $m$}
\label{alg:combin}
\begin{algorithmic}
\Procedure{Local search}{}
 \State Let $D_0$ be size $m$ design 
 \State Set $\delta = 1$ and $D \leftarrow D_0$ 
 \While{$\delta > 0$}
 \ForAll{element $\mathcal{E}_k \in D$ and $\mathcal{E}_{k'}\in \mathcal{D} / D$}
    Calculate $f_c(D / \{\mathcal{E}_k\} \cup \{\mathcal{E}_{k'}\})$ 
 \EndFor
 \State Set $D' \leftarrow \argmin_{k,k'} f_c(D / \{\mathcal{E}_k\} \cup \{\mathcal{E}_{k'}\})$ 
 \State $\delta = f_c(D') - f_c(D)$ 
 \If{$\delta > 0$ }
    $D \leftarrow D'$
 \EndIf
 \EndWhile
\EndProcedure


 \Procedure{Reverse greedy search}{}
\State $D \gets \mathcal{D}$
 \State $l \gets N$\;
 \While{$l > m$}
 \ForAll{$\mathcal{E}_{k}\in D$}
    \State Calculate $f_c(D / \{\mathcal{E}_{k}'\})$
 \EndFor
 \State Set $D \leftarrow D / \argmin_{\mathcal{E}_k} f_c(D / \{\mathcal{E}_k\})$ \;
 \State $l \leftarrow l - 1$
 \EndWhile
 \EndProcedure
\end{algorithmic}
 \end{algorithm}

The local search algorithm starts from a design of the desired size $m$ and then makes the swap of an experimental unit in the design with one not in the design that leads to the greatest reduction in the c-optimality criterion. Such swaps are made until no further value improving swaps are available. The worst possible design that this algorithm produces under a cardinality constraint (i.e. $\vert d \vert \leq J$) has a value no larger than $3/2$ times the true c-optimal design.\cite{Fisher1978} This bound can be improved to $1 + 1/e$ with certain extensions to the algorithm.\cite{Filmus2014} 

The reverse greedy algorithm starts from the complete design space and successively removes the experimental unit that results in the largest decrease in variance.  Proofs of the constant factor approximation for the reverse greedy algorithm depends on the `steepness' or `curvature' of $f(.)$, which depends on the value of $f(\emptyset)$, where $\emptyset$ is the empty set, i.e. a design with no observations. A reasonable choice for the variance of an estimator from a design with no observations is infinity, as we specify in (\ref{eq:coptim}). However, the resulting curvature of the function then means there is no constant factor approximation bound.\cite{Ilev2001,Sviridenko2017} Alternatively we could say $f(\emptyset)$ is undefined, and we would again lack a theoretical guarantee.


Watson and Pan\cite{Watson2022} investigate these algorithms for a range of study designs, including cluster randomised trials. They find that empirically the reverse greedy and local search algorithms provide similar performance in terms of the variance of the resulting design. The reverse greedy search is deterministic, while the local search starts from a random design, so Watson and Pan run the local search multiple times and select the best design. They also suggest several approaches to improve the computational efficiency of these algorithms.

Kasza and Forbes\cite{Kasza2019} use a reverse greedy approach to identify optimal designs. They describe the method as estimating the `information content' of clusters or cluster-periods in a design space like Figure \ref{fig:desspace}, where their measure of information is the marginal change in variance from removing the observations from the design. The results presented by Kasza and Forbes\cite{Kasza2019} are qualitatively similar to those using other methods and algorithms, such as those presented below. 

Hooper, Kasza, and Forbes (2020)\cite{Hooper2020} examine optimal cluster trial designs in the context of the linear mixed model with covariance function AR1. They consider a discrete approximation to a continuous time model with continuous recruitment and polynomial functions of time. The design space consists of individuals regularly spaced over a time interval within clusters; the individuals constitute the experimental unit. They aim to provide a set of illustrative optimal designs under different parameter values for the covariance function. The method used to identify these designs could also be described as a variant of the `reverse greedy' algorithm. Each iteration of the algorithm is supplemented with a type of local search, although the swaps of experimental units that can be made are limited at each step to preserve a no reversibility restriction. The designs presented by Hooper, Kasza, and Forbes\cite{Hooper2021} are often qualitatively different from those presented here resulting from other methods. However, the design space they use includes a wide range of other designs, and their specfication of $X$ does not include time period indiciators, which may account for some of the differences. 

\subsubsection{Computational Complexity}
The computational complexity of the local and greedy searches scales as $O(m^4r^3(J-m))$ and $O(J^3r^3(J-m)$, respectively,\cite{Watson2022} where $r$ is the number of observations in an experimental unit. These algorithms scale relatively poorly with the size of the design. However, the approach taken by Girling and Hemming\cite{Girling2016} discussed above suggests a way of improving the computational time of these algorithms when the experimental unit is a cluster or cluster-period. Equation (\ref{eq:lmm3}) specifies a model for the cluster-period mean under covariance function EXC2. A similar model can be specified for the AR1 function with equal sized cluster-periods:
 \begin{align}
 \begin{split}
 \label{eq:lmm4}
    \Bar{y}_{kt} &= J_{kt}\delta + W_t\tau_t + \alpha_{kt} + e_{kt} \\
    \text{Cov}(\alpha_{kt},\alpha_{kt'}) &= \tau^2 \lambda^{\vert t - t' \vert} \\
    \text{Var}(e_{kt}) &= \frac{\sigma^2}{r}
    \end{split}
\end{align}
The advantage of using a model for the cluster-periods is that it only requires a single swap or addition to change an experimental unit as opposed to $n$ swaps or additions to the design. 

\subsection{Non-Gaussian Models}
The multiplicative weighting, optimal mixed model weights, and combinatorial methods all require calculation of the covariance matrix $\Sigma$ and its inverse. For Gaussian models with identity link function $\Sigma = \sigma^2I + ZDZ^T$, so it can be calculated exactly. For non-Gaussian models, such as Binomial or Poisson, generating $\Sigma$ can be computationally demanding. For non-linear models, an approximation to $\Sigma$ and hence to the information matrix $M$, is typically used.\cite{Waite2015} Breslow and Clayton (1993)\cite{Breslow1993} used the marginal quasilikelihood of the GLMM to propose the first-order approximation:
\begin{equation}
\label{eq:sigapprox}
    \Sigma \approx W^{-1} + ZDZ^T
\end{equation}
where $W$ is a diagonal matrix with entries $W_{i,i} = \left(\left(\frac{\partial \mu}{\partial \eta} \right)^2 \text{Var}(y|\mathbf{u}) \right)$, which are the GLM iterated weights.\cite{mccullagh2019generalized} Here, $W$ is evaluated at the marginal mean $X\beta$. For the optimal mixed model weights algorithm we can generate $\Sigma = \frac{1}{N}W^{-1}\text{diag}(\boldsymbol{\phi}^{-1}) + ZDZ^T$.

Zeger et al (1988)\cite{Zeger1988} suggest that when using the marginal quasilikelihood, approximations can be improved by `attenuating' the linear predictor. For example, for the binomial-logit model one would use $\mu_i = h^{-1}(x_i\beta\vert aDz_i^Tz_i + I\vert^{-1/2})$ where $a = 16\sqrt{3}/15\pi$. For other types of optimality this attenuation can improve the resulting designs,\cite{Waite2015} however for \textit{c}-optimality there was little evidence of a difference in the designs considered by Watson and Pan.\cite{Watson2022} Other information matrix approximations that may be relevant for non-Gaussian models include using the GEE working covariance matrix or higher order approximations, however, these methods are either more restrictive or there is little evidence they improve the designs. The approximation also permits the use of cluster-period mean models, like (\ref{eq:lmm3}) and (\ref{eq:lmm4}), with heteroskedastic errors given by $\text{Var}(e_{jt}) = \frac{W_{jt,jt}}{r_{jt}}$ where the $r_{jt}$ is the number of observations in cluster $j$ at time period $t$, and $W_{jt,jt}$ the individual-level variance of an observation in that cluster-period. For the non-Gaussian examples we give below, we use Equation \ref{eq:sigapprox} without attenutation. 

Morbeek and Maas (2005)\cite{Moerbeek2005} examine optimal designs for clustered studies with a binomial-logisitic mixed model. The derive an approximation to the variance of the treatment effect parameter under the EXC1 covariance function using a linearisation approach with the marginal quasilikelihood. They specifically aim to identify the optimal number of individuals within a cluster in a cost-benefit framework.

\subsection{Robust Optimality}
The methods to generate an optimal design have so far assumed the model parameters are known. However, a well known issue for optimal experimental design methodology is that a design that may be optimal for one set of parameters or model specification may perform poorly for another. Robust methods that are efficient across a range of designs are therefore desirable. There are multiple possible criteria for modifying the c-optimal design criterion to account for multiple designs. For example, Girling and Hemming\cite{Girling2016} consider a minimax criterion in which they identify a design that maximises (minimises) the minimum (maximum) precision (variance) over all values of the correlation between cluster-period means. This results in a `hybrid' trial design (see Figure \ref{fig:desspace}). Van Breukelen and Candel (2015)\cite{VanBreukelen2015} also consider a minimax criterion to identify a robust optimal cluster trial design when the ICC is unknown. Similarly to Moerbeek\cite{Moerbeek2005}, they use a cost-benefit framework and examine the optimal design under a fixed budget.

As a robust optimality criterion, the maximin function is not necessarily generally applicable. For the combinatorial methods, we require that the objective function is supermodular, and the maximum of a set of supermodular functions is not necessarily supermodular. As an alternative, we can use a `weighted average'. In particular, we assume there is a set of $L$ candidate models and we specify a prior probability for each model $p_1,...,p_L$ with the property $\sum_{l=1}^L p_l = 1$. Dette (1993)\cite{Dette1993} and Lauter (1974)\cite{Lauter1974} propose the following generalisation of the c-optimality criterion:
\begin{equation}
\label{eq:extcoptim}
    f(d;\mathcal{A}) = \sum_{l=1}^L p_l \log(c_l^T M^{-1}_{d,l} c^T_l)
\end{equation}
where $M_{d,l} = (X^T_{d,(l)}\Sigma^{-1}_{d,(l)}X_{d,(l)})^{-1}$ represents the information matrix for design $D$ under the $l$th model. As well as the parameters varying between model specification, the vectors $c_{l}$ and matrices $X_{(l)}$ and $\Sigma_{(l)}$ can vary between models, for example, there may be different specifications of time and covariance functions. 

Dette\cite{Dette1993} generalises the Elfving theorem for this robust criterion for models with uncorrelated observations. One can further generalise this theorem to the case where observations are correlated within experimental units following the results of Holland-Letz et al\cite{Holland-Letz2011} and Sagnol\cite{Sagnol2011}. However, a specification for a program to solve this generalised problem using conic optimisation methods, extending the results of Sagnol in the single model case, is not currently available, and remains a topic for future research. An extension of the optimal mixed model weights method to robust optimal designs is similarly an open question.

Another robust c-optimality criterion is the weighted average:
\begin{equation}
\label{eq:extcoptim2}
    f(d;\mathcal{A}) = \sum_{l=1}^L p_l c_l^T M^{-1}_{d,l} c^T_l
\end{equation}
Both this criterion and (\ref{eq:extcoptim}) can be used with the combinatorial search methods, since they are also supermodular and maintain the same theoretical guarantees. Following Dette\cite{Dette1993} we describe a design that maximises either of these criteria as being c-optimal for the class $\mathcal{A}$ \textit{with respect to} the prior $p$. 

\subsection{Code examples}
We have provided code samples and examples using the \texttt{glmmrOptim} package, including code to reproduce the figures in this article at \url{https://samuel-watson.github.io/glmmr-web/other/optimal_examples/}.

\section{Results and Examples}
In this section we provide a range of examples to illustrate the use of the methods and summarise results from several of the papers cited above. For the combinatorial algorithms, we use the reverse greedy algorithm. For multiplicative weighting methods, we select the best design from a variety of different rounding methods. Where applicable we also compare the results to those presented by Girling and Hemming\cite{Girling2016}.

\subsection{Clusters as Experimental Units}
For the first set of examples we consider Design Space A in Figure \ref{fig:desspace} with seven unique cluster sequences and six time periods. Our goal is to identify a design of $m=10$ clusters. Each row is repeated up to five times in the design space, which is to say each sequence could be duplicated up to five times in the final design. Limiting the number of duplicate sequences to five, rather than ten, prevents the final design being, for example, a purely before and after design, while permitting parallel, stepped-wedge, and hybrid designs (although, we have not found a scenario where before-and-after design is optimal). Before and after designs may not be desirable as they lack any randomised comparison; treatment status will be correlated strongly with secular temporal trends, which is why they are unlikely to be optimal. We consider the linear mixed models given in Equations (\ref{eq:lmm3}) and (\ref{eq:lmm4}) with EXC2 and AR1 covariance functions, respectively. The method proposed by Girling and Hemming is applicable in the EXC2 case (the scenario here is the same as that given in Figure 5 of Girling and Hemming\cite{Girling2016}).

\begin{figure}
    \centering
    \includegraphics[width=\textwidth]{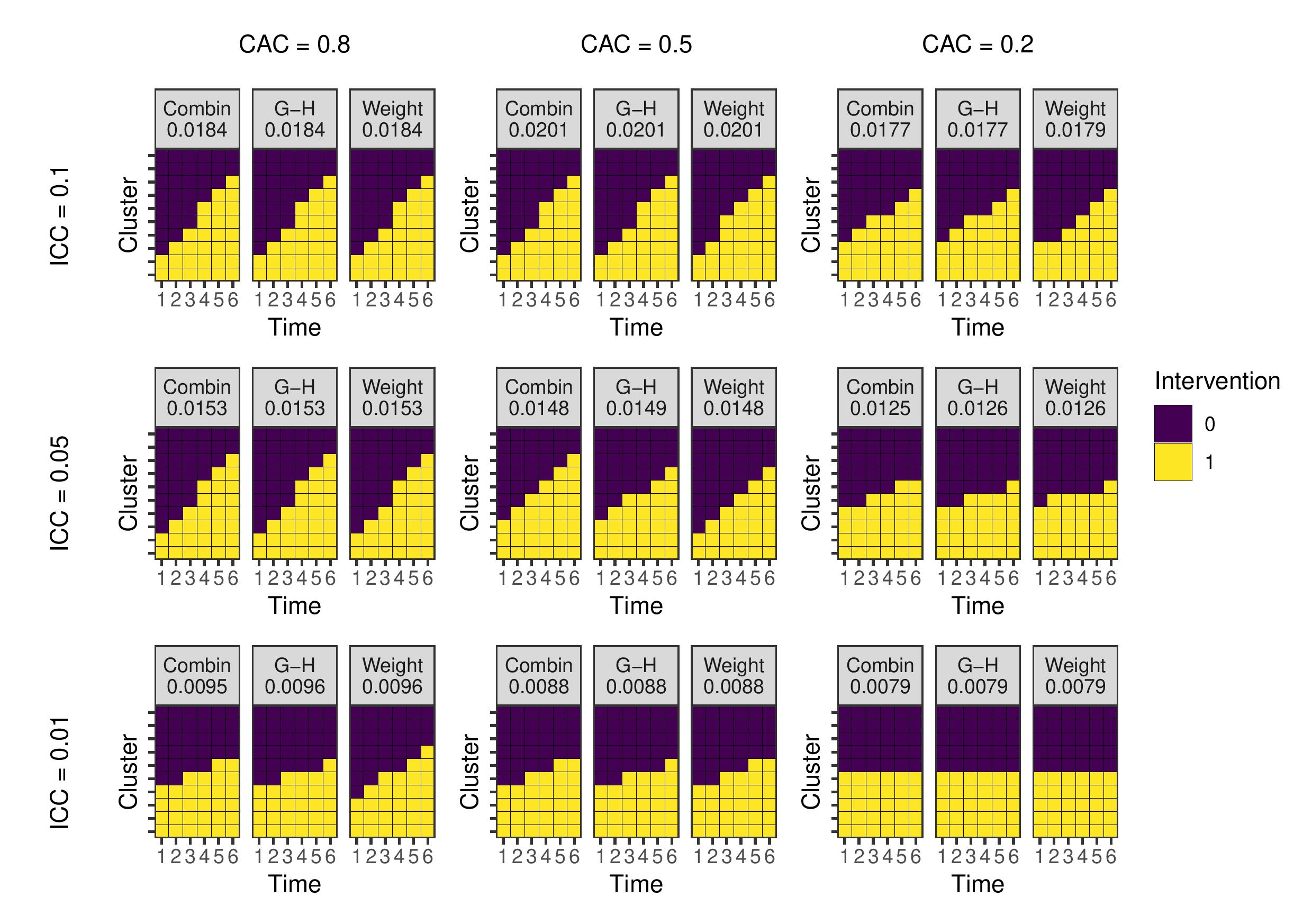}
    \caption{Optimal study designs with ten clusters and six time periods for different values of the ICC and CAC using a linear mixed model with EXC2 covariance structure with $m=10$ individuals per cluster-period. `Combin' are results from the combinatorial local search run 100 times and selecting the best design, `G-H' are results using the method from Girling and Hemming, and `Weight' are designs produced by estimating experimental unit weights. The number is the estimator variance from the design.}
    \label{fig:ex1}
\end{figure}

\begin{figure}
    \centering
    \includegraphics[width=\textwidth]{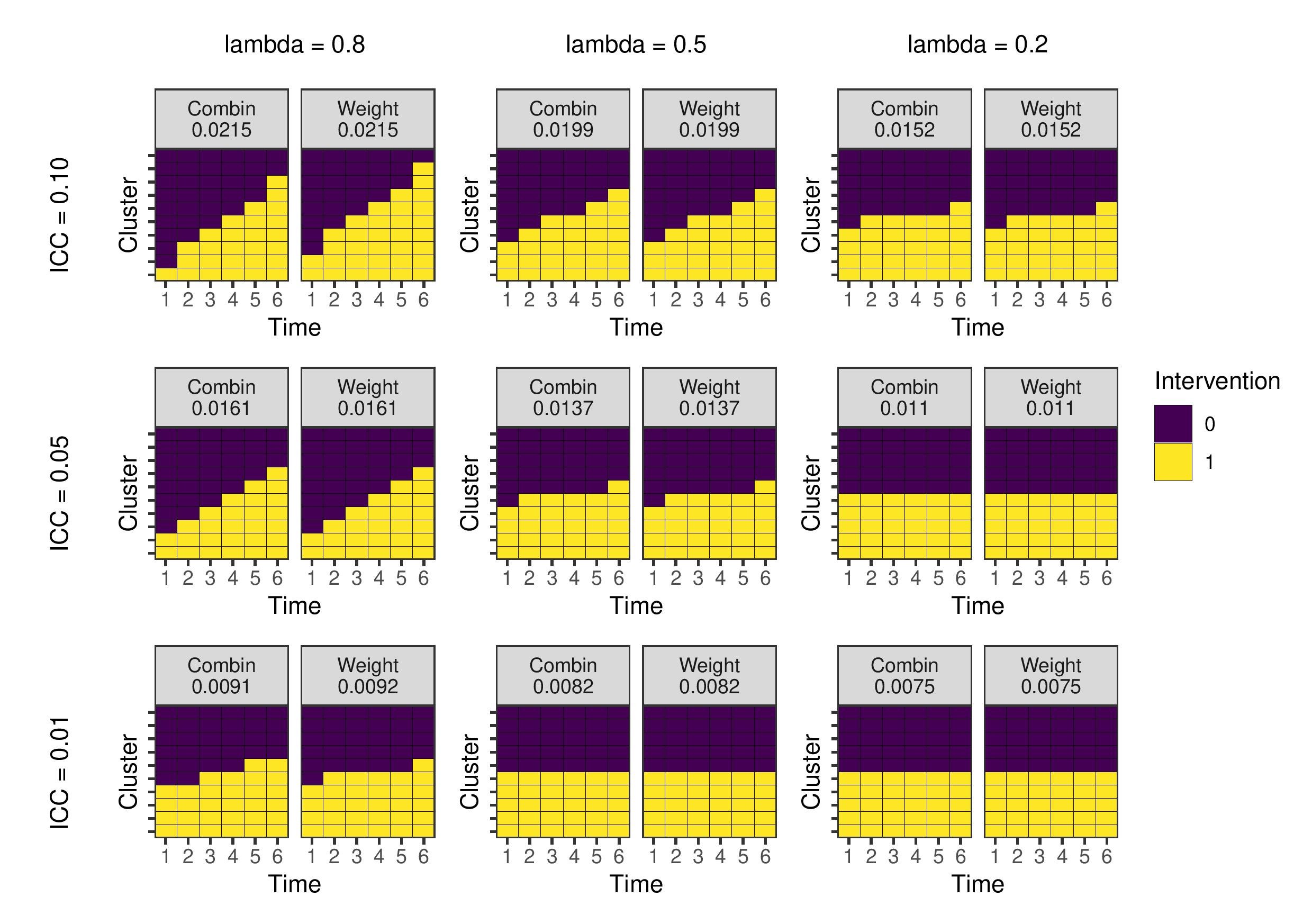}
    \caption{Optimal study designs with ten clusters and six time periods for different values of the ICC and autoregressive parameter $\lambda$ (`lambda') using a linear mixed model with AR1 covariance structure with $m=10$ individuals per cluster-period. `Combin' are results from the combinatorial local search run 100 times and selecting the best design and `Weight' are designs produced by estimating experimental unit weights. The number is the estimator variance from the design.}
    \label{fig:ex2}
\end{figure}

Figure \ref{fig:ex1} shows the results using the EXC2 covariance function with $m=10$ individuals per cluster-period. The resulting designs for each set of covariance parameters are the same from each method, with only a couple of exceptions. However, the difference between the variances from the designs do not exceed 0.0001. In all cases, the design from the combinatorial method has the lowest variance. Figure \ref{fig:ex2} shows the results from the model with AR1 covariance function. As with EXC2, the designs are generally the same from both combinatorial and weighting methods, but where there is a difference, the combinatorial method produces a design with marginally lower variance. For both covariance functions, as the level of correlation within a cluster and between periods or the overall level of within cluster-period correlation gets higher, the degree of `staggering' increases.

\begin{figure}
    \centering
    \begin{subfigure}[b]{\textwidth}
        \centering
        \includegraphics[width=\textwidth]{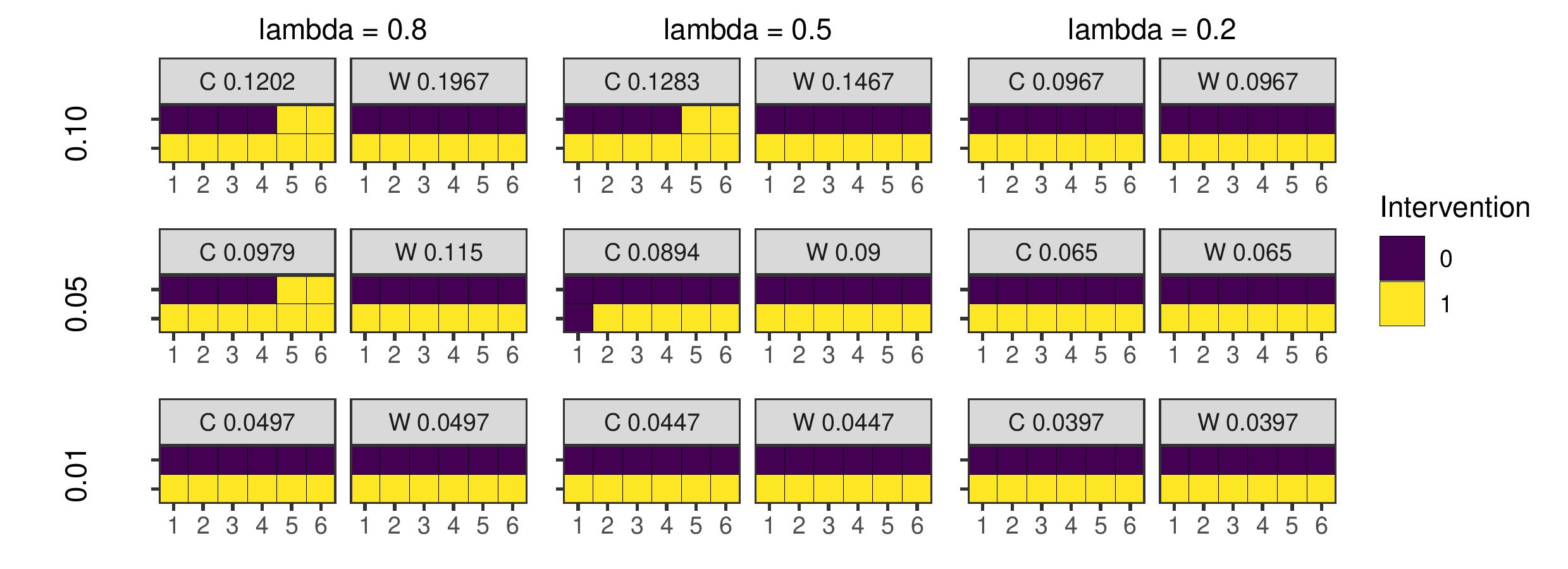}
        \caption{EXC2 covariance function}
    \end{subfigure}
    \hfill
    \begin{subfigure}[b]{\textwidth}
        \centering
        \includegraphics[width=\textwidth]{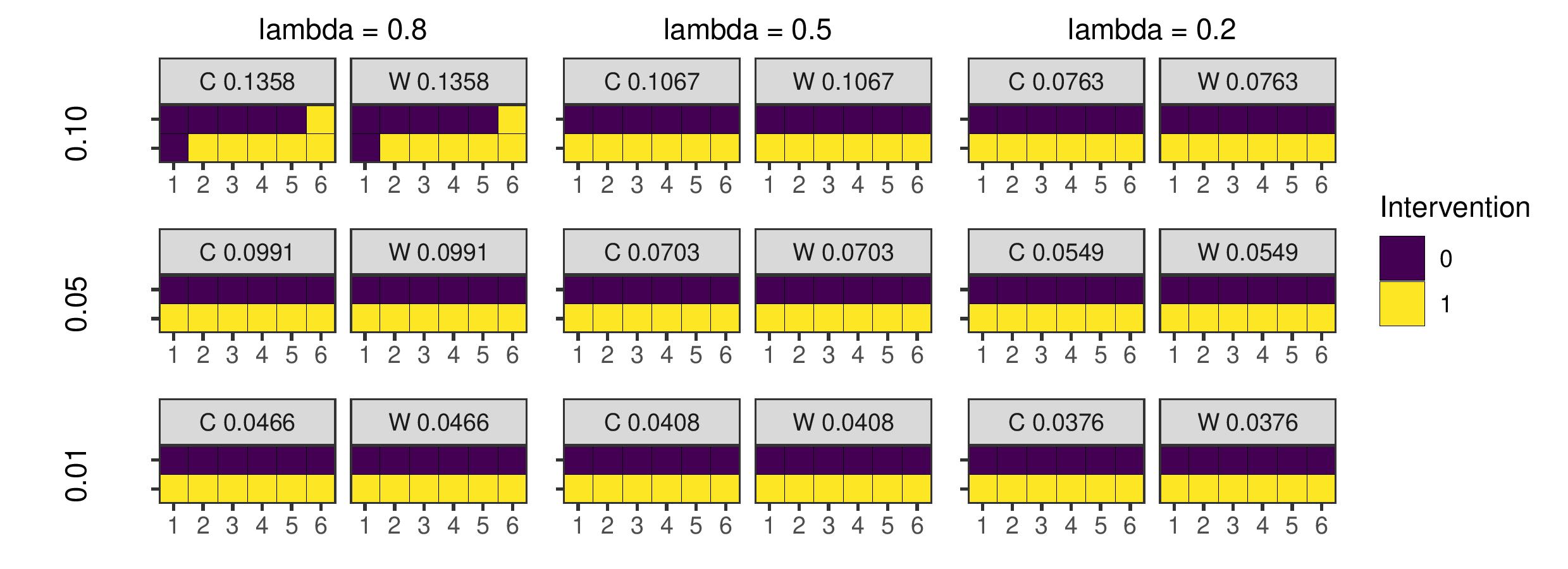}
        \caption{AR1 covariance function}
    \end{subfigure}
    \caption{Optimal study designs of two cluster sequences and six time periods for different values of the covariance parameters with the EXC2 and AR1 covariance functions. C = Combinatorial local search. W = experimental unit weights. The number on each panel is the treatment effect estimator variance for the design. The rows are difference values of the ICC.}
    \label{fig:extwop}
\end{figure}

The previous example assumes any design might be permissible within the design space. However, more restrictive design problems may be of interest given practical limitations on intervention roll out. As an example, we may require there to be only two trial arms within which all clusters receive the intervention at the same time. The question is then when each arm should receive the intervention (if at all). We can consider this problem as selecting two experimental units from Design Space A containing the seven experimental units in Figure \ref{fig:desspace}, since the variance of this design is proportional to a design with $J$ clusters allocated 1:1 to each of the two sequences. Figure \ref{fig:extwop} shows the optimal two cluster sequences using combinatorial and weighting methods. The two methods agree for all parameter values with the AR1 covariance function, however, for the EXC2 function the weighting method produces designs with higher variance. For low values of the CAC or $\lambda$ and the ICC a parallel design is optimal. For higher values of these parameters, inclusion of baseline or endline observations in which both trial arms are in control or treatment states, respectively, is superior to a purely parallel design.

\subsection{Single Observations as Experimental Units}

\begin{figure}
    \centering
    \includegraphics[width=\textwidth]{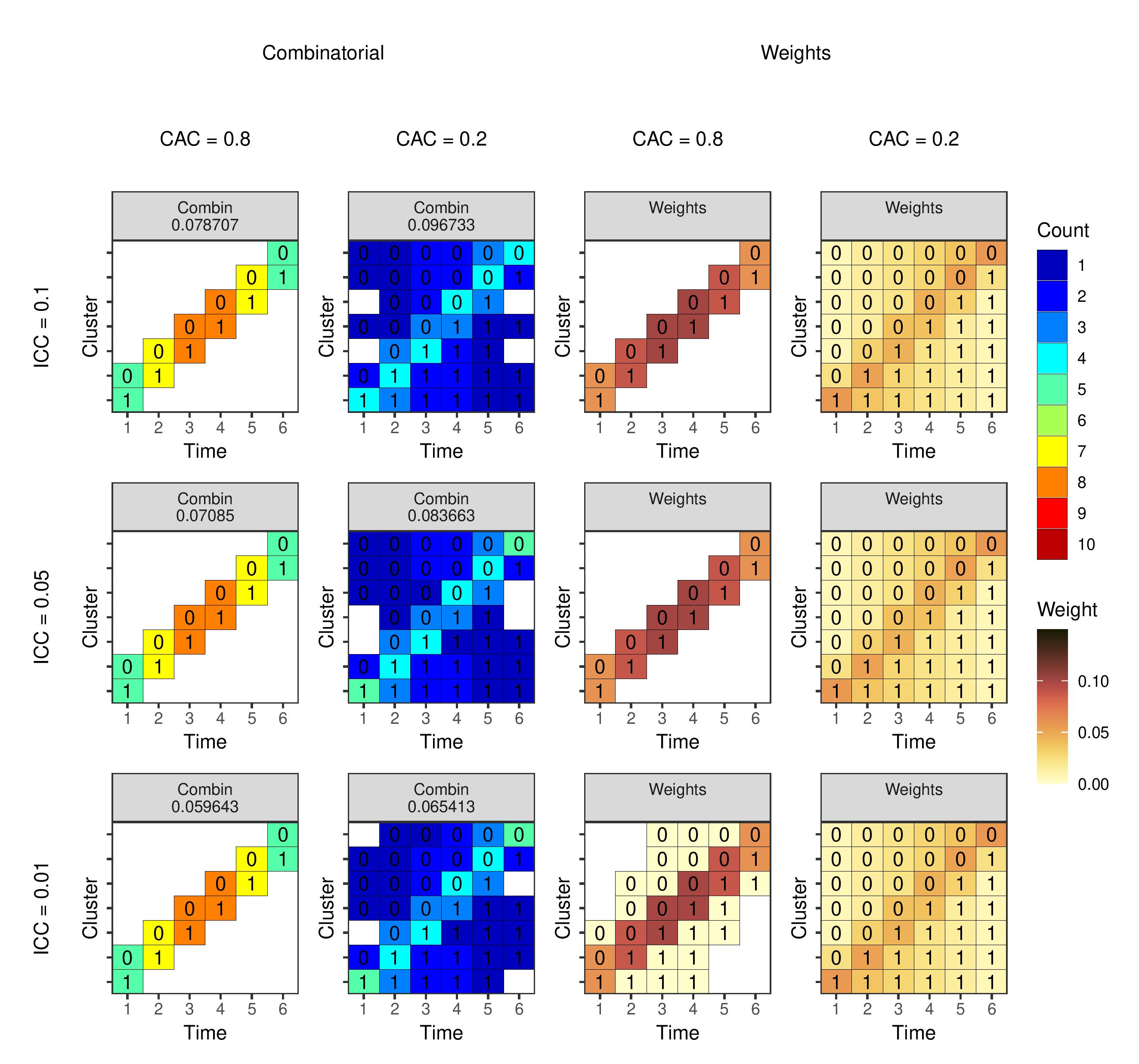}
    \caption{Optimal study designs of 80 individuals with seven clusters and six time periods using a linear mixed model with EXC2 covariance structure with different values of the ICC (rows) and CAC (columns). Results from the combinatorial reverse greedy search (with up to ten individuals per cluster-period) and optimal mixed model weights algorithms. The number for the left two columns is the estimator variance from the design. The number within each cell is the intervention status and the colour represents the number of observations (left two columns) or the weight (right two columns).}
    \label{fig:ex3}
\end{figure}

\begin{figure}
    \centering
    \includegraphics[width=\textwidth]{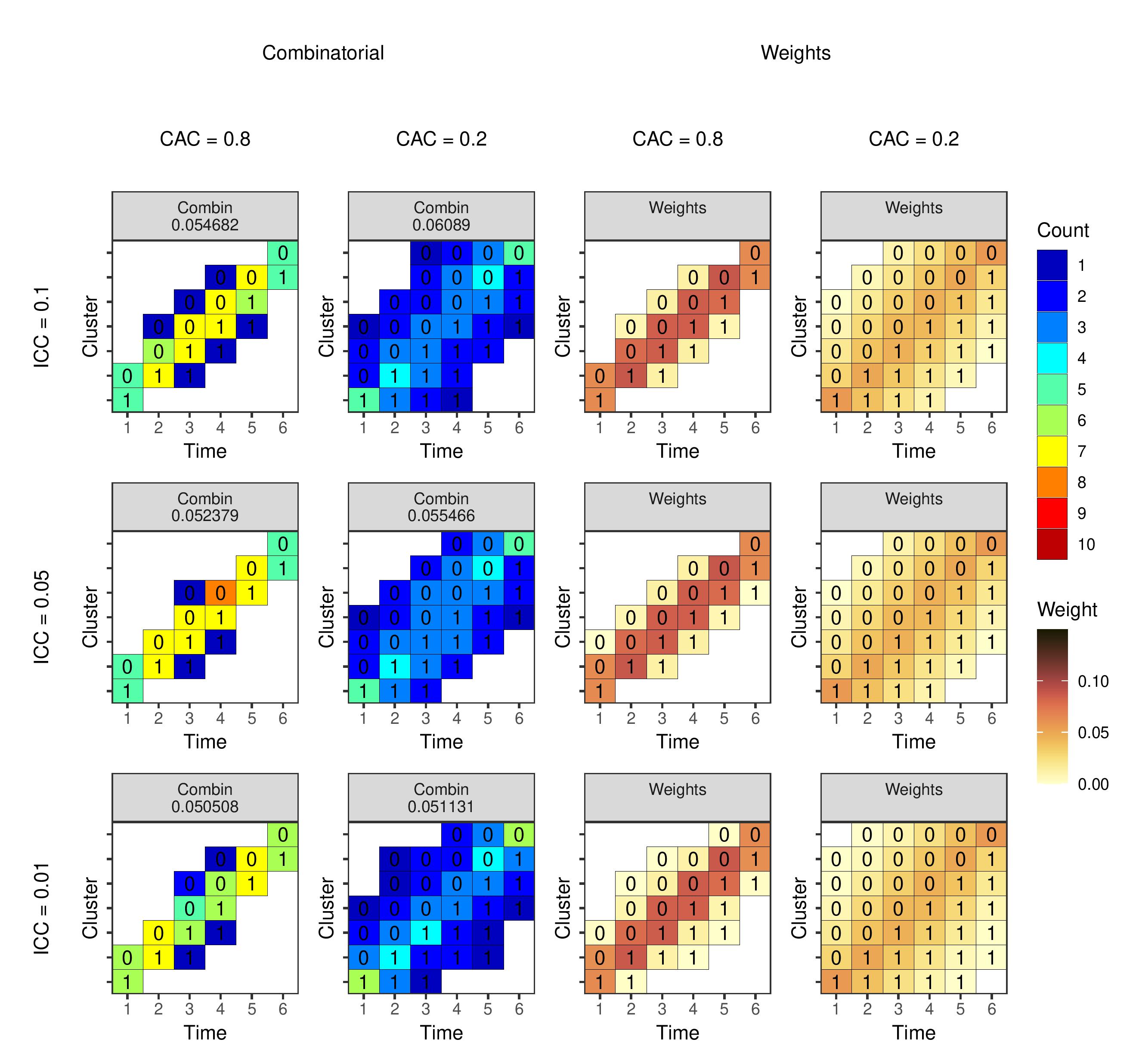}
    \caption{Optimal study designs of 80 individuals for different values of the ICC and $\lambda$ using a linear mixed model with AR1 covariance structure with different values of the ICC (rows) and autoregressive parameter $\lambda$ (columns). Results from the combinatorial reverse greedy search (with up to ten individuals per cluster-period) and optimal mixed model weights algorithms. The number for the left two columns is the estimator variance from the design. The number within each cell is the intervention status and the colour represents the number of observations (left two columns) or the weight (right two columns).}
    \label{fig:ex4}
\end{figure}

For the next examples we specify a single observation as the experimental unit. The design space is as specified in Figure \ref{fig:desspace} with seven clusters and six time periods, and each cluster-period has ten unique individuals who each contribute an observation. Using the combinatorial algorithms, our goal here is to select 80 observations of the 420 possible observations up to a maximum of ten per cluster-period. The mixed model weights can also be calculated using Algorithm \ref{alg:girling} for comparison. Figures \ref{fig:ex3} and \ref{fig:ex4} show the results for the EXC2 and AR1 covariance functions, respectively. In general, the levels of within cluster-period correlation (CAC or $\lambda$) appear to determine the optimal design, with higher levels resulting in greater numbers of observations placed along the main diagonal. Not all the designs are exactly symmetric, which may suggest the algorithm has not found the exactly optimal design.

\subsection{Non-Gaussian Models}
For non-Gaussian models, we illustrate how the parameters $\beta$ affect the resulting optimal design. We consider the design problem given for the examples shown in Figures \ref{fig:ex3} and \ref{fig:ex4} with single observations as experimental units and Design Space A of Figure \ref{fig:desspace} with up to ten individuals per cluster-period. We specify a binomial-logistic model. In all the examples we use parameters $\tau^2 = 0.16$ and $\omega^2 = 0.04$ for EXC2 or $\tau^2 = 0.20$ and $\lambda = 0.8$ for AR1, giving an approximate ICC of 0.05. The time period parameters are specified to give a control group mean outcome proportion of either 5\%, 25\%, or 50\% and odds ratios for the six time periods of 0.8, 0.9, 1.0, 1.0, 1.1, and 1.2, respectively. The treatment effect is an odds ratio of either 0.5 or 1.5. 

\begin{figure}
    \centering
    \includegraphics[width=\textwidth]{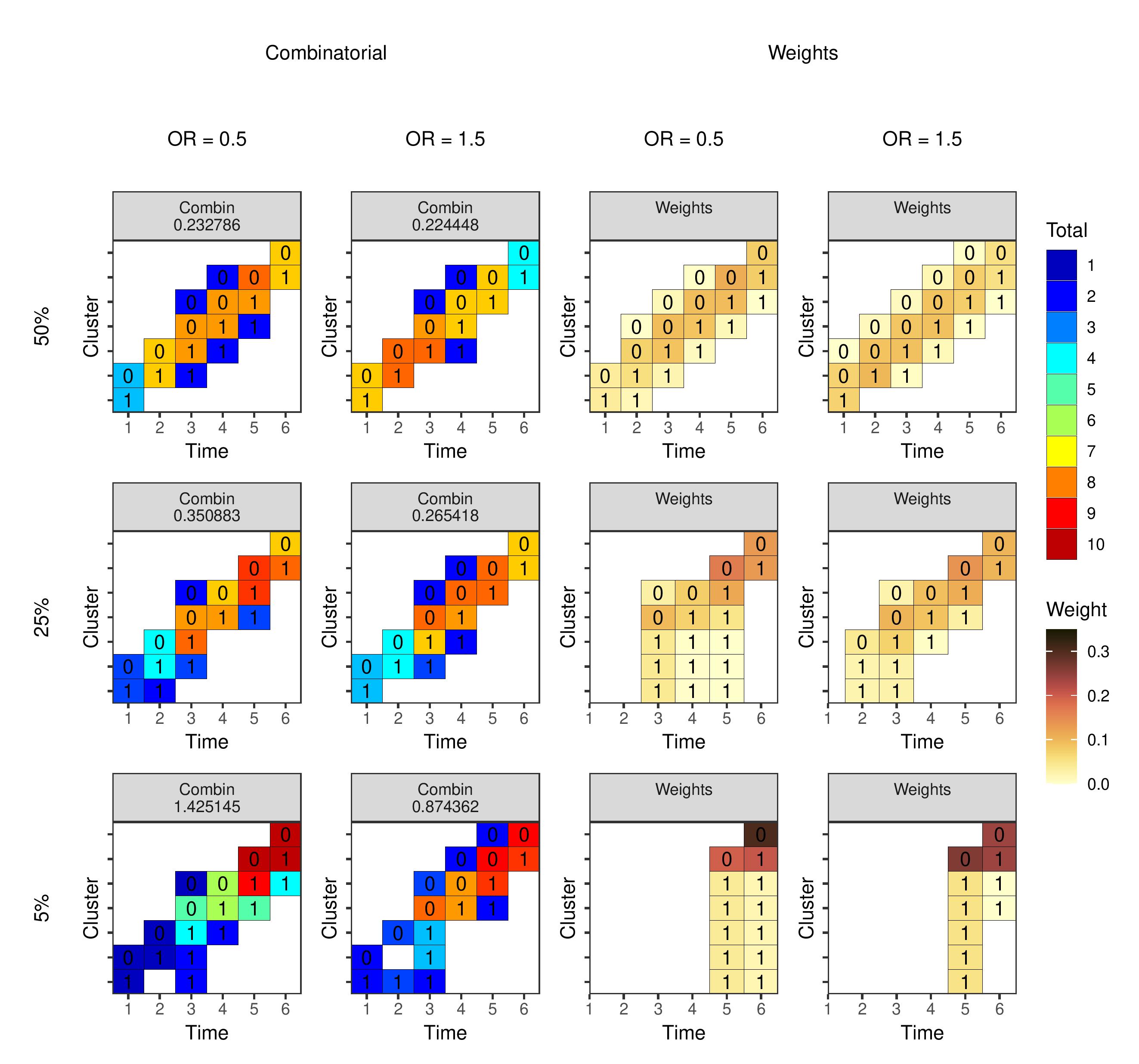}
    \caption{Optimal study designs of 80 indivudals with ten clusters and six time periods for different values of the base rate (rows) and intervention effect size (columns) with a binomial-logistic mixed model. Results from the combinatorial reverse greedy search (with up to ten individuals per cluster-period) and optimal mixed model weights algorithms. The number for the left two columns is the estimator variance from the design. The number within each cell is the intervention status and the colour represents the number of observations (left two columns) or the weight (right two columns).}
    \label{fig:ex5}
\end{figure}

Figure \ref{fig:ex5} shows the optimal designs of 80 individuals for the binomial-logistic example using the combinatorial and optimal mixed model weight algorithms. When the base rate is low, the relative difference in individual-level variance between time periods is larger, and the resulting designs favour placing more observations in those later time periods. When the base rate is higher, the designs more closely resemble those from the linear model in Figures \ref{fig:ex3} and \ref{fig:ex4}. The optimal weights suggest that when the base rate is low in this example, we should place all our efforts in the last periods; the combinatorial algorithms have specified a cap of ten observations per cluster-period and so distribute the observations in the next-best cluster-periods.

\subsection{Robust Optimal Designs}
\begin{figure}
    \centering
    \includegraphics[width=\textwidth]{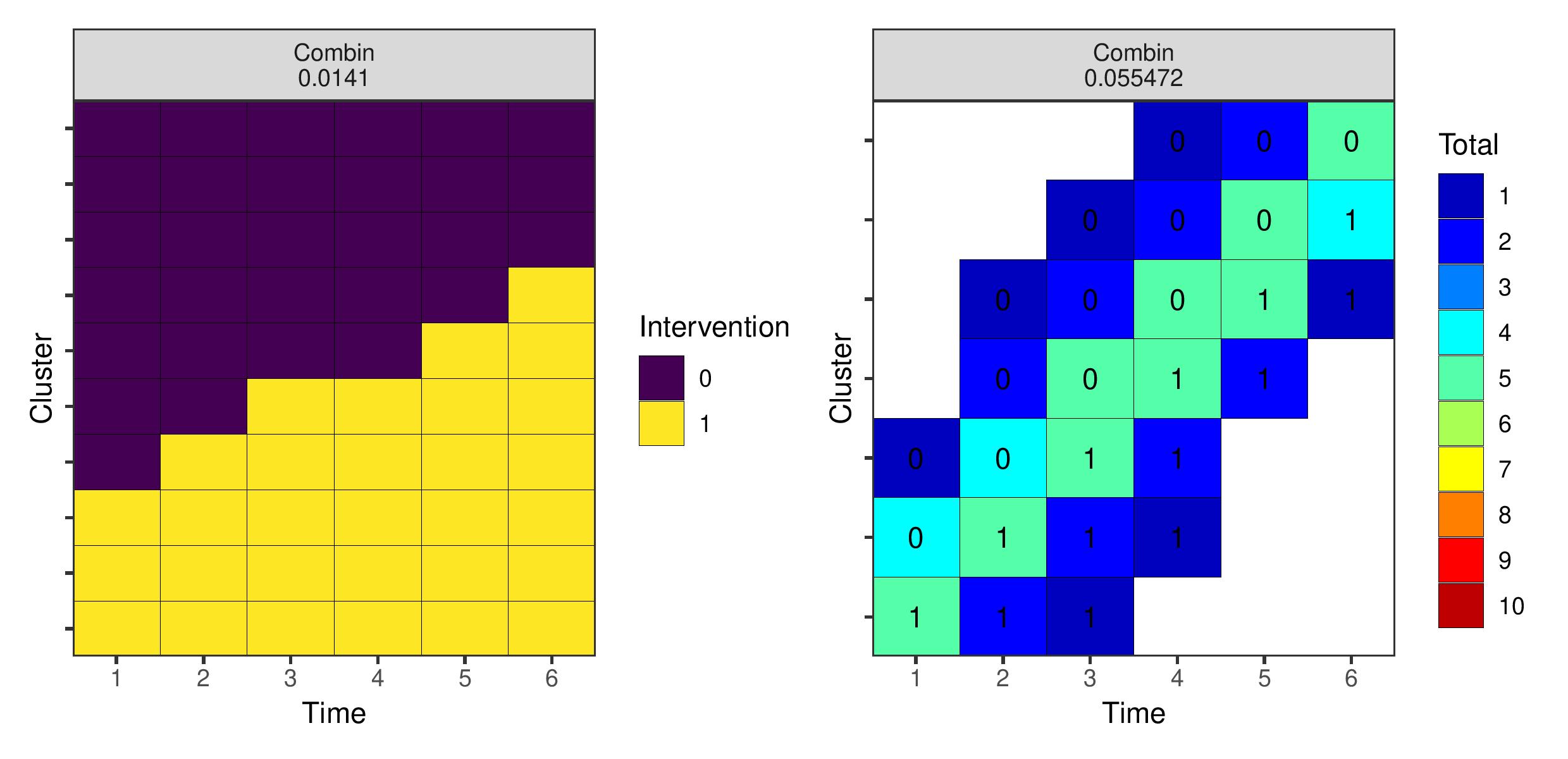}
    \caption{Robust optimal study designs of 80 indivudals with ten clusters and six time periods with respect to a prior that weights each possibility from earlier examples equally. Results from the combinatorial local search run 100 times and selecting the best design. The left panel is for a design space with clusters as experimental units, and the right panel where individuals are experimental units. The numbers in the cells on the right panel show the intervention status.}
    \label{fig:ex6}
\end{figure}

To illustrate robust optimal designs we consider the 18 models and parameter values represented by the panels Figure 1 and 2. We assume that there is no prior knowledge of the likely values of the covariance parameters, nor the covariance function, and so assign equal prior weights to all 18 designs. We use the weighted average robust criterion (\ref{eq:extcoptim2}), and run the local search algorithm 100 times, selecting the lowest variance design. The left panel of Figure \ref{fig:ex6} shows the resulting optimal design with respect to the equal weighting prior. Similarly to Girling and Hemming\cite{Girling2016}, the design is a `hybrid' trial design with six of ten clusters following a parallel trial design, and the remaining four a staggered implementation roll-out. We also identify a robust optimal design for individual experimental units with the 18 designs shown in Figures \ref{fig:ex3} and \ref{fig:ex4} using the same procedure. The resulting design is shown in the right-hand panel of Figure \ref{fig:ex6}. 

\section{Discussion and Conclusions}
\subsection{Comparison of algorithms}
The correlation between observations in a cluster randomised trial setting complicates identification of optimal study designs. Indeed, there have been relatively few studies on the topic of optimal cluster trial designs, particularly when compared with individual-level randomised controlled trials. However, recent methodological advances provide several approaches for approximating c-optimal designs with correlated observations. 

We have discussed three different types of method within a general framework for cluster trials with discrete time: using exact formulae for specific models specifications and design spaces and using an algorithm or enumerating and evaluating multiple relevant designs; determining weights to place on each experimental units in a design space; and, combinatorial algorithms for selecting an optimal subset of experimental units. These categories are not exhaustive and new methods may be developed using novel approaches. Each of the three types of method has their advantages and disadvantages. Minimising exact functions for the estimator variance would be preferable, but explicit formulae are only available in the simpler cases. Many authors (e.g.\cite{Girling2016,Zhan2018,Lawrie2015}) consider the linear mixed model with cluster and cluster-period exchangeable random effects, for example.  The combinatorial algorithms produced the lowest variance design in all the examples we considered where we could compare methods, but were generally more computationally demanding when one takes into account the suggestion to run the algorithm multiple times and select the best design. The optimal mixed model weights algorithm identifies the optimal weights for each cluster-period, although may not produce an exact design when rounding the totals. The optimal mixed model weights algorithm is much faster to run than other generic algorithms. For the examples presented in Figures \ref{fig:ex3} and \ref{fig:ex4}, the reverse greedy search took around one minute, the local search ten seconds, and the model weights 50 milliseconds. In many circumstances, it is difficult or impractical to specify exact numbers of individuals, and so weights would be sufficient, in which case the mixed model weights are likely the best choice given its efficiency. However, for more complex design problems, such as setting maximum or minimum numbers of observations in different cluster-periods, the combinatorial approaches may be required. 

\subsection{Small sample bias}
A well recognised issue for cluster trials, and GLMMs in general, is that the generalised least squares estimator of the standard errors of $\beta$ in Equation (\ref{eq:infomat}) exhibits a small sample bias. The standard errors for $\hat{\beta}$ are underestimated when the number of clusters is small (see, e.g. \cite{Leyrat2018,Kahan2016,Watson2021b}). All of the examples given in this article may well suffer from this issue. There are two reasons for the bias. First, the information matrix $M_d$ is estimated in practice by evaluating the the covariance matrix at the estimated values of the covariance parameters. The GLS estimator (\ref{eq:infomat}) does not account from this additional variability from estimating the covariance parameters. Second, the estimator for the information matrix is itself a biased estimator the variance of $\hat{\beta}$. Kackar and Harwell\cite{Kackar1984} describe an approximation to the small sample variance of $\hat{\beta}$ for linear mixed models that accounts for the estimation of the covariance parameters and Kenward and Rogers\cite{Kenward1997} extend this approximation to also account for the bias. One might consider therefore using these ``corrected'' estimators in place of the generalised least squares information matrix in the optimality criterion. However, it is not clear whether this approach would perform well or not; both corrections are first-order approximations and can exhibit behaviour that may undermine the performance of the algorithms. For example, in exploratory testing we found it was possible, while using fixed covariance parameter values, for a smaller design to have a marginally smaller ``corrected'' variance than a larger design. The algorithms produced similar, but not identical, ``optimal'' designs using these corrected matrices though. Optimal designs with small sample corrections thus remains an important topic for future research in this area. 

\subsection{Usefulness of optimal designs}
Optimal designs are not always practical. For example, many of the designs in Figures \ref{fig:ex3} to \ref{fig:ex6} where the experimental unit was the individual included cluster-periods with a single individual. It is very unlikely that this would ever be implementable in practice given the logistics of data collection within clusters such as hospitals, clinics, or schools. However, one can view these optimal designs as a benchmark against which to justify a chosen study design. Hooper\cite{Hooper2021} suggests that there is a common misconception among cluster trial practitioners that the stepped-wedge design is more efficient than a parallel trial. The results of Girling and Hemming,\cite{Girling2016} which are replicated in Figure \ref{fig:ex1}, and others show that this is not the case.  The most efficient design depends on the covariance parameters, and in the case of a non-linear model, the parameters in the linear predictor too. Indeed, a useful heuristic is that emerges from these results is that the less variable the cluster means over time, the more `variable' the intervention should be (i.e. more staggered over time). Identifying an optimal design can help design a practicable trial that is more efficient than might otherwise be considered. Where individual-level experimental units are used, it can identify which cluster-periods to exclude entirely and which to place more effort into. Kasza et al\cite{Kasza2019} propose just such an approach based on a `reverse greedy' type algorithm. 

The framework we use to present these methods requires enumeration of all the unique experimental units. For more complex design problems the design space can then become very large. For example, Hooper et al\cite{Hooper2021b} use a discrete approximation to a continuous time model, and aim to identify when a cluster should start and stop recruiting and when it should implement the intervention. There is a very large number of possible cluster sequences that would fit within this design space given the large number of time increments, even with the no reversibility and symmetric restrictions they use. Enumerating the complete design space and subjecting it to one of the algorithms above would likely be highly computationally demanding. Indeed, this issue raises the question of how one might approach cluster trial optimal design question with continuous time. Other examples in the literature in which a treatment variable is potentially continuous, have relied on selecting a small number of discrete possible values; the finer the discretisation the better the result.\cite{Yang2013} Extending this to larger numbers of possible conditions, or treating time as truly continuous thus remains a topic of future research. One potentially useful approach for continuous time models may be `particle swarm' optimization and other `nature inspired' methods.\cite{Chen2022}

\subsection{Bayesian optimal designs}
We have also not considered Bayesian optimal design. While Bayesian methods are relatively rarely used for the design and analysis of cluster randomised trials, there are growing number of examples (e.g.\cite{ThriveatWorkWellbeingProgrammeCollaboration2019}). Chaloner\cite{Chaloner1995} provides a review of Bayesian optimal experimental design criteria. Bayesian optimal designs are based on maximising a utility function for the experiment. The resulting optimality criteria though are highly similar to their Frequentist counterparts, but they introduce the added complexity of needing to integrate over the prior distributions of the model parameters. There have been several methodological advances and new algorithms proposed for identifying Bayesian optimal experiemental designs. For example, Overstall and Woods (2017)\cite{Overstall2017} provide perhaps the most general solution to this problem for non-linear Bayesian models. The algorithms in this article might also be used to find approximate solutions to Bayesian cluster trial design problems. For example, the robust criterion (\ref{eq:extcoptim2}) could be translated to a Bayesian context where the weights are derived using a Riemann sum approximation to the integral over the prior distributions.\cite{Watson2022} However, further research is required into Bayesian methods for the design and analysis of cluster randomised trials.

\subsection{Conclusion}
The final choice of study design for a cluster randomised trial results from the confluence of a range of practical, financial, and statistical considerations. However, there is an ethical obligation to try to minimise the sample size required to achieve a research objective. Methods to identify optimal or approximately optimal study designs therefore serve a useful purpose where there is flexibility in the roll out of an intervention. We have identified several methods relevant to cluster randomised trials, which can be used on a standard computer in a short amount of time. We would therefore suggest that examining the optimal trial design should be a step in the design of every cluster randomised trial.







\begin{funding}
This work was supported with funding from the Medical Research Council MR/V038591/1.
\end{funding}

\bibliographystyle{SageV}
\bibliography{main}

\begin{thebibliography}{10}
\providecommand{\url}[1]{\texttt{#1}}
\providecommand{\urlprefix}{URL }
\expandafter\ifx\csname urlstyle\endcsname\relax
  \providecommand{\doi}[1]{DOI:\discretionary{}{}{}#1}\else
  \providecommand{\doi}{DOI:\discretionary{}{}{}\begingroup
  \urlstyle{rm}\Url}\fi
\providecommand{\eprint}[2][]{\url{#2}}

\bibitem{Eldridge2012}
Eldridge S and Kerry S.
\newblock \emph{{A Practical Guide to Cluster Randomised Trials in Health
  Services Research}}.
\newblock Chichester, UK: John Wiley {\&} Sons, Ltd, 2012.
\newblock ISBN 9781119966241.
\newblock \doi{10.1002/9781119966241}.
\newblock \urlprefix\url{http://doi.wiley.com/10.1002/9781119966241}.

\bibitem{Murray1998}
Murray DM.
\newblock \emph{{Design and Analysis of Group Randomised Trials}}.
\newblock New York, NY: Oxford University Press Inc., 1998.

\bibitem{Hooper2016}
Hooper R, Teerenstra S, de~Hoop E et~al.
\newblock {Sample size calculation for stepped wedge and other longitudinal
  cluster randomised trials}.
\newblock \emph{Statistics in Medicine} 2016; 35(26): 4718--4728.
\newblock \doi{10.1002/sim.7028}.
\newblock \urlprefix\url{http://doi.wiley.com/10.1002/sim.7028}.

\bibitem{Hemming2020}
Hemming K, Kasza J, Hooper R et~al.
\newblock {A tutorial on sample size calculation for multiple-period cluster
  randomized parallel, cross-over and stepped-wedge trials using the Shiny CRT
  Calculator}.
\newblock \emph{International Journal of Epidemiology} 2020; dyz237.
\newblock \doi{10.1093/ije/dyz237}.
\newblock
  \urlprefix\url{https://academic.oup.com/ije/advance-article/doi/10.1093/ije/dyz237/5748155}.

\bibitem{Li2021}
Li F, Hughes JP, Hemming K et~al.
\newblock {Mixed-effects models for the design and analysis of stepped wedge
  cluster randomized trials: An overview}.
\newblock \emph{Statistical Methods in Medical Research} 2021; 30(2): 612--639.
\newblock \doi{10.1177/0962280220932962}.
\newblock
  \urlprefix\url{http://journals.sagepub.com/doi/10.1177/0962280220932962}.

\bibitem{Atkinson2007}
Atkinson AC, Donev AN and Tobias RD.
\newblock \emph{Optimum Experimental Design, with SAS}.
\newblock Clarendon, 2007.

\bibitem{Berger2009}
Berger MPF and Wong WK.
\newblock \emph{An Introduction to Optimal Designs for Social and Biomedical
  Research}.
\newblock Wiley, 2009.

\bibitem{Yang2013}
Yang M, Biedermann S and Tang E.
\newblock {On Optimal Designs for Nonlinear Models: A General and Efficient
  Algorithm}.
\newblock \emph{Journal of the American Statistical Association} 2013;
  108(504): 1411--1420.
\newblock \doi{10.1080/01621459.2013.806268}.
\newblock
  \urlprefix\url{http://www.tandfonline.com/doi/abs/10.1080/01621459.2013.806268}.

\bibitem{Girling2016}
Girling AJ and Hemming K.
\newblock {Statistical efficiency and optimal design for stepped cluster
  studies under linear mixed effects models}.
\newblock \emph{Statistics in Medicine} 2016; 35(13): 2149--2166.
\newblock \doi{10.1002/sim.6850}.
\newblock \urlprefix\url{https://onlinelibrary.wiley.com/doi/10.1002/sim.6850}.

\bibitem{Hooper2021}
Hooper R and Eldridge SM.
\newblock {Cutting edge or blunt instrument: how to decide if a stepped wedge
  design is right for you}.
\newblock \emph{BMJ Quality {\&} Safety} 2021; 30(3): 245--250.
\newblock \doi{10.1136/bmjqs-2020-011620}.
\newblock
  \urlprefix\url{https://qualitysafety.bmj.com/lookup/doi/10.1136/bmjqs-2020-011620}.

\bibitem{Watson2021}
Watson SI, Girling AJ and Hemming K.
\newblock {Design and analysis of three-arm parallel group randomised trials
  with small numbers of clusters}.
\newblock \emph{Statistics in Medicine} 2021; 40(5): 1133--1146.

\bibitem{Hussey2007}
Hussey MA and Hughes JP.
\newblock {Design and analysis of stepped wedge cluster randomized trials}.
\newblock \emph{Contemporary Clinical Trials} 2007; 28(2): 182--191.
\newblock \doi{10.1016/j.cct.2006.05.007}.
\newblock
  \urlprefix\url{https://linkinghub.elsevier.com/retrieve/pii/S1551714406000632}.

\bibitem{Lawrie2015}
Lawrie J, Carlin JB and Forbes AB.
\newblock {Optimal stepped wedge designs}.
\newblock \emph{Statistics {\&} Probability Letters} 2015; 99: 210--214.
\newblock \doi{10.1016/j.spl.2015.01.024}.
\newblock
  \urlprefix\url{https://linkinghub.elsevier.com/retrieve/pii/S0167715215000309}.

\bibitem{Woertman2013}
Woertman W, de~Hoop E, Moerbeek M et~al.
\newblock Stepped wedge designs could reduce the required sample size in
  cluster randomized trials.
\newblock \emph{Journal of Clinical Epidemiology} 2013; 66: 752--758.
\newblock \doi{10.1016/j.jclinepi.2013.01.009}.

\bibitem{Zhan2018}
Zhan Z, de~Bock GH and van~den Heuvel ER.
\newblock {Optimal unidirectional switch designs}.
\newblock \emph{Statistics in Medicine} 2018; 37(25): 3573--3588.
\newblock \doi{10.1002/sim.7853}.
\newblock \urlprefix\url{https://onlinelibrary.wiley.com/doi/10.1002/sim.7853}.

\bibitem{Hooper2021b}
Hooper R and Copas AJ.
\newblock {Optimal design of cluster randomised trials with continuous
  recruitment and prospective baseline period}.
\newblock \emph{Clinical Trials} 2021; 18(2): 147--157.
\newblock \doi{10.1177/1740774520976564}.
\newblock
  \urlprefix\url{http://journals.sagepub.com/doi/10.1177/1740774520976564}.

\bibitem{Copas2020}
Copas AJ and Hooper R.
\newblock {Cluster randomised trials with different numbers of measurements at
  baseline and endline: Sample size and optimal allocation}.
\newblock \emph{Clinical Trials} 2020; 17(1): 69--76.
\newblock \doi{10.1177/1740774519873888}.
\newblock
  \urlprefix\url{http://journals.sagepub.com/doi/10.1177/1740774519873888}.

\bibitem{Moerbeek2020}
Moerbeek M.
\newblock {Optimal designs for group randomized trials and group administered
  treatments with outcomes at the subject and group level}.
\newblock \emph{Statistical Methods in Medical Research} 2020; 29(3): 797--810.
\newblock \doi{10.1177/0962280219846149}.
\newblock
  \urlprefix\url{http://journals.sagepub.com/doi/10.1177/0962280219846149}.

\bibitem{Lemme2018}
Lemme F, van Breukelen GJ and Candel MJ.
\newblock Efficient treatment allocation in 2 × 2 multicenter trials when
  costs and variances are heterogeneous.
\newblock \emph{Statistics in Medicine} 2018; 37: 12--27.
\newblock \doi{10.1002/sim.7499}.

\bibitem{Elfving1952}
Elfving G.
\newblock {Optimum Allocation in Linear Regression Theory}.
\newblock \emph{The Annals of Mathematical Statistics} 1952; 23(2): 255--262.
\newblock \doi{10.1214/aoms/1177729442}.
\newblock \urlprefix\url{http://projecteuclid.org/euclid.aoms/1177729442}.

\bibitem{Holland-Letz2011}
Holland-Letz T, Dette H and Pepelyshev A.
\newblock {A geometric characterization of optimal designs for regression
  models with correlated observations}.
\newblock \emph{Journal of the Royal Statistical Society: Series B (Statistical
  Methodology)} 2011; 73(2): 239--252.
\newblock \doi{10.1111/j.1467-9868.2010.00757.x}.
\newblock
  \urlprefix\url{https://onlinelibrary.wiley.com/doi/10.1111/j.1467-9868.2010.00757.x}.

\bibitem{Sagnol2011}
Sagnol G.
\newblock {Computing optimal designs of multiresponse experiments reduces to
  second-order cone programming}.
\newblock \emph{Journal of Statistical Planning and Inference} 2011; 141(5):
  1684--1708.
\newblock \doi{10.1016/j.jspi.2010.11.031}.
\newblock
  \urlprefix\url{https://linkinghub.elsevier.com/retrieve/pii/S0378375810005318}.

\bibitem{HollandLetz2012}
Holland-Letz T, Dette H and Renard D.
\newblock Efficient algorithms for optimal designs with correlated observations
  in pharmacokinetics and dose-finding studies.
\newblock \emph{Biometrics} 2012; 68: 138--145.
\newblock \doi{10.1111/j.1541-0420.2011.01657.x}.

\bibitem{Balinski2002}
Balinski M and Young P.
\newblock \emph{{Fair Representation: Meeting the Ideal of One Man, One Vote}}.
\newblock 2nd editio ed. Washington D.C.: Brookings Institution Press, 2002.
\newblock ISBN 0-8157-0111-X.

\bibitem{PUKELSHEIM1992}
PUKELSHEIM F and RIEDER S.
\newblock {Efficient rounding of approximate designs}.
\newblock \emph{Biometrika} 1992; 79(4): 763--770.
\newblock \doi{10.1093/biomet/79.4.763}.
\newblock
  \urlprefix\url{https://academic.oup.com/biomet/article-lookup/doi/10.1093/biomet/79.4.763}.

\bibitem{Fedorov1972}
Fedorov V.
\newblock \emph{{Theory of Optimal Experiments}}.
\newblock New York: Academic Press, 1972.

\bibitem{Hemming2015}
Hemming K, Haines TP, Chilton PJ et~al.
\newblock {The stepped wedge cluster randomised trial: rationale, design,
  analysis, and reporting}.
\newblock \emph{BMJ} 2015; 350(feb06 1): h391--h391.
\newblock \doi{10.1136/bmj.h391}.
\newblock \urlprefix\url{http://www.bmj.com/cgi/doi/10.1136/bmj.h391}.

\bibitem{Pukelsheim2006}
Pukelsheim F.
\newblock \emph{{Optimal Design of Experiments}}.
\newblock 2nd editio ed. Society for Industrial and Applied Mathematics, 2006.
\newblock ISBN 978-0-89871-604-7.
\newblock \doi{10.1137/1.9780898719109}.
\newblock
  \urlprefix\url{http://epubs.siam.org/doi/book/10.1137/1.9780898719109}.

\bibitem{Watson2022}
Watson SI and Pan Y.
\newblock {Approximate c-Optimal Experimental Designs with Correlated
  Observations using Combinatorial Optimisation}.
\newblock \emph{Statistics and Computing} 2022;
  \urlprefix\url{http://arxiv.org/abs/2207.09183}.
\newblock \eprint{(In press)}.

\bibitem{Sviridenko2017}
Sviridenko M, Vondr{\'{a}}k J and Ward J.
\newblock {Optimal Approximation for Submodular and Supermodular Optimization
  with Bounded Curvature}.
\newblock \emph{Mathematics of Operations Research} 2017; 42(4): 1197--1218.
\newblock \doi{10.1287/moor.2016.0842}.
\newblock
  \urlprefix\url{http://pubsonline.informs.org/doi/10.1287/moor.2016.0842}.

\bibitem{Wynn1970}
Wynn HP.
\newblock {The Sequential Generation of {\$}D{\$}-Optimum Experimental
  Designs}.
\newblock \emph{The Annals of Mathematical Statistics} 1970; 41(5): 1655--1664.
\newblock \doi{10.1214/aoms/1177696809}.
\newblock \urlprefix\url{http://projecteuclid.org/euclid.aoms/1177696809}.

\bibitem{Fisher1978}
Fisher ML, Nemhauser GL and Wolsey LA.
\newblock {An analysis of approximations for maximizing submodular set
  functions—II}.
\newblock \emph{Mathematical Programming} 1978; 15: 265--294.
\newblock \doi{10.1007/bfb0121195}.

\bibitem{Nemhauser1978}
Nemhauser GL and Wolsey LA.
\newblock {Best Algorithms for Approximating the Maximum of a Submodular Set
  Function}.
\newblock \emph{Mathematics of Operations Research} 1978; 3(3): 177--188.
\newblock \doi{10.1287/moor.3.3.177}.
\newblock
  \urlprefix\url{http://pubsonline.informs.org/doi/abs/10.1287/moor.3.3.177}.

\bibitem{Filmus2014}
Filmus Y and Ward J.
\newblock {Monotone Submodular Maximization over a Matroid via Non-Oblivious
  Local Search}.
\newblock \emph{SIAM Journal on Computing} 2014; 43(2): 514--542.
\newblock \doi{10.1137/130920277}.
\newblock \urlprefix\url{http://epubs.siam.org/doi/10.1137/130920277}.

\bibitem{Ilev2001}
Il'ev VP.
\newblock {An approximation guarantee of the greedy descent algorithm for
  minimizing a supermodular set function}.
\newblock \emph{Discrete Applied Mathematics} 2001; 114(1-3): 131--146.
\newblock \doi{10.1016/S0166-218X(00)00366-8}.
\newblock
  \urlprefix\url{https://linkinghub.elsevier.com/retrieve/pii/S0166218X00003668}.

\bibitem{Kasza2019}
Kasza J and Forbes AB.
\newblock {Information content of cluster–period cells in stepped wedge
  trials}.
\newblock \emph{Biometrics} 2019; 75(1): 144--152.
\newblock \doi{10.1111/biom.12959}.
\newblock
  \urlprefix\url{https://onlinelibrary.wiley.com/doi/10.1111/biom.12959}.

\bibitem{Hooper2020}
Hooper R, Kasza J and Forbes A.
\newblock {The hunt for efficient, incomplete designs for stepped wedge trials
  with continuous recruitment and continuous outcome measures}.
\newblock \emph{BMC Medical Research Methodology} 2020; 20(1): 279.
\newblock \doi{10.1186/s12874-020-01155-z}.
\newblock
  \urlprefix\url{https://bmcmedresmethodol.biomedcentral.com/articles/10.1186/s12874-020-01155-z}.

\bibitem{Waite2015}
Waite TW and Woods DC.
\newblock {Designs for generalized linear models with random block effects via
  information matrix approximations}.
\newblock \emph{Biometrika} 2015; 102(3): 677--693.
\newblock \doi{10.1093/biomet/asv005}.

\bibitem{Breslow1993}
Breslow NE and Clayton DG.
\newblock {Approximate Inference in Generalized Linear Mixed Models}.
\newblock \emph{Journal of the American Statistical Association} 1993; 88(421):
  9--25.
\newblock \doi{10.1080/01621459.1993.10594284}.
\newblock
  \urlprefix\url{https://www.tandfonline.com/doi/full/10.1080/01621459.1993.10594284}.

\bibitem{mccullagh2019generalized}
McCullagh P and Nelder JA.
\newblock \emph{Generalized linear models, 2nd Edition}.
\newblock Routledge, 1989.

\bibitem{Zeger1988}
Zeger SL, Liang KY and Albert PS.
\newblock {Models for Longitudinal Data: A Generalized Estimating Equation
  Approach}.
\newblock \emph{Biometrics} 1988; 44(4): 1049--1060.
\newblock \urlprefix\url{https://www.jstor.org/stable/2531734}.

\bibitem{Moerbeek2005}
Moerbeek M and Maas CJM.
\newblock {Optimal Experimental Designs for Multilevel Logistic Models with Two
  Binary Predictors}.
\newblock \emph{Communications in Statistics - Theory and Methods} 2005; 34(5):
  1151--1167.
\newblock \doi{10.1081/STA-200056839}.
\newblock
  \urlprefix\url{http://www.tandfonline.com/doi/abs/10.1081/STA-200056839}.

\bibitem{VanBreukelen2015}
van Breukelen GJ and Candel MJ.
\newblock {Efficient design of cluster randomized and multicentre trials with
  unknown intraclass correlation}.
\newblock \emph{Statistical Methods in Medical Research} 2015; 24(5): 540--556.
\newblock \doi{10.1177/0962280211421344}.
\newblock
  \urlprefix\url{http://journals.sagepub.com/doi/10.1177/0962280211421344}.

\bibitem{Dette1993}
Dette H.
\newblock {Elfving's Theorem for {\$}D{\$}-Optimality}.
\newblock \emph{The Annals of Statistics} 1993; 21(2).
\newblock \doi{10.1214/aos/1176349149}.
\newblock
  \urlprefix\url{https://projecteuclid.org/journals/annals-of-statistics/volume-21/issue-2/Elfvings-Theorem-for-D-Optimality/10.1214/aos/1176349149.full}.

\bibitem{Lauter1974}
L{\"{a}}uter E.
\newblock {Experimental design in a class of models}.
\newblock \emph{Mathematische Operationsforschung und Statistik} 1974; 5(4-5):
  379--398.
\newblock \doi{10.1080/02331887408801175}.
\newblock
  \urlprefix\url{https://www.tandfonline.com/doi/full/10.1080/02331887408801175}.

\bibitem{Leyrat2018}
Leyrat C, Morgan KE, Leurent B et~al.
\newblock {Cluster randomized trials with a small number of clusters: Which
  analyses should be used?}
\newblock \emph{International Journal of Epidemiology} 2018; 47(1): 321--331.
\newblock \doi{10.1093/ije/dyx169}.

\bibitem{Kahan2016}
Kahan BC, Forbes G, Ali Y et~al.
\newblock {Increased risk of type I errors in cluster randomised trials with
  small or medium numbers of clusters: A review, reanalysis, and simulation
  study}.
\newblock \emph{Trials} 2016; \doi{10.1186/s13063-016-1571-2}.

\bibitem{Watson2021b}
Watson SI, Girling AJ and Hemming K.
\newblock {Design and analysis of three-arm parallel group randomised trials
  with small numbers of clusters}.
\newblock \emph{Statistics in Medicine} 2021; 40(5): 1133--1146.

\bibitem{Kackar1984}
Kackar RN and Harville DA.
\newblock {Approximations for Standard Errors of Estimators of Fixed and Random
  Effects in Mixed Linear Models}.
\newblock \emph{Journal of the American Statistical Association} 1984; 79(388):
  853--862.
\newblock \doi{10.1080/01621459.1984.10477102}.
\newblock
  \urlprefix\url{http://www.tandfonline.com/doi/abs/10.1080/01621459.1984.10477102}.

\bibitem{Kenward1997}
Kenward MG and Roger JH.
\newblock {Small Sample Inference for Fixed Effects from Restricted Maximum
  Likelihood}.
\newblock \emph{Biometrics} 1997; 53(3): 983.
\newblock \doi{10.2307/2533558}.
\newblock \urlprefix\url{https://www.jstor.org/stable/2533558?origin=crossref}.

\bibitem{Chen2022}
Chen P, Chen R and Wong WK.
\newblock Particle swarm optimization for searching efficient experimental
  designs: A review.
\newblock \emph{WIREs Computational Statistics} 2022; 14.
\newblock \doi{10.1002/wics.1578}.

\bibitem{ThriveatWorkWellbeingProgrammeCollaboration2019}
at~Work Wellbeing Programme~Collaboration T.
\newblock Evaluation of a policy intervention to promote the health and
  wellbeing of workers in small and medium sized enterprises - a cluster
  randomised controlled trial.
\newblock \emph{BMC public health} 2019; 19: 493.
\newblock \doi{10.1186/s12889-019-6582-y}.
\newblock
  \urlprefix\url{https://bmcpublichealth.biomedcentral.com/articles/10.1186/s12889-019-6582-y
  http://www.ncbi.nlm.nih.gov/pubmed/31046713
  http://www.pubmedcentral.nih.gov/articlerender.fcgi?artid=PMC6498586}.

\bibitem{Chaloner1995}
Chaloner K and Verdinelli I.
\newblock Bayesian experimental design: A review.
\newblock \emph{Statistical Science} 1995; 10.
\newblock \doi{10.1214/ss/1177009939}.

\bibitem{Overstall2017}
Overstall AM and Woods DC.
\newblock {Bayesian Design of Experiments Using Approximate Coordinate
  Exchange}.
\newblock \emph{Technometrics} 2017; 59(4): 458--470.
\newblock \doi{10.1080/00401706.2016.1251495}.
\newblock
  \urlprefix\url{https://www.tandfonline.com/doi/full/10.1080/00401706.2016.1251495}.

\end{thebibliography}
\clearpage
\appendix
\section{Additional Results}
\begin{figure}
    \centering
    \includegraphics[width=\textwidth]{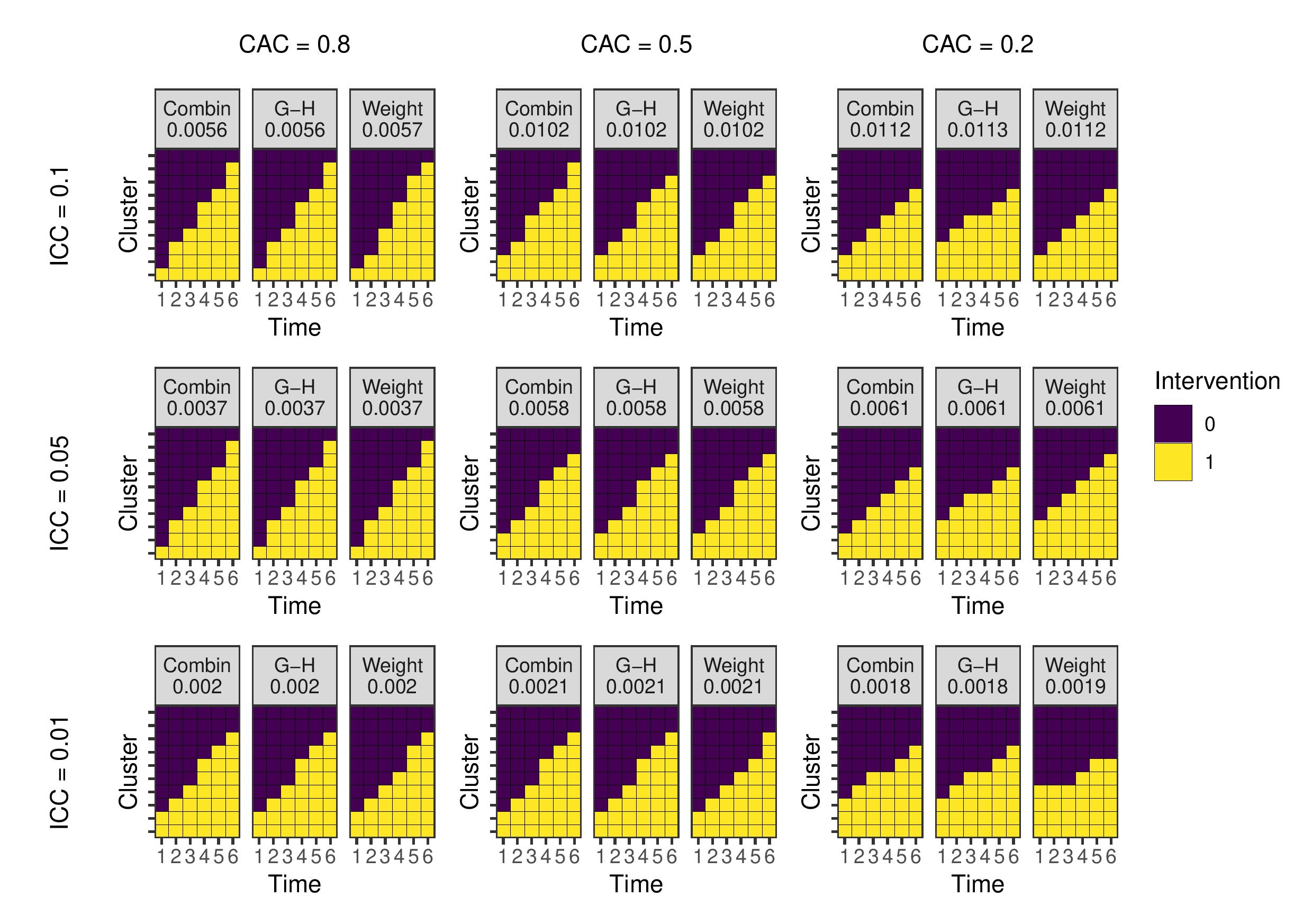}
    \caption{Optimal study designs with ten clusters and six time periods for different values of the ICC and CAC using a linear mixed model with EXC2 covariance structure with $m=100$ individuals per cluster-period. `Combin' are results from the combinatorial local search run 100 times and selecting the best design, `G-H' are results using the method from Girling and Hemming, and `Weight' are designs produced by estimating experimental unit weights.}
    \label{fig:ex1b}
\end{figure}

\begin{figure}
    \centering
    \includegraphics[width=\textwidth]{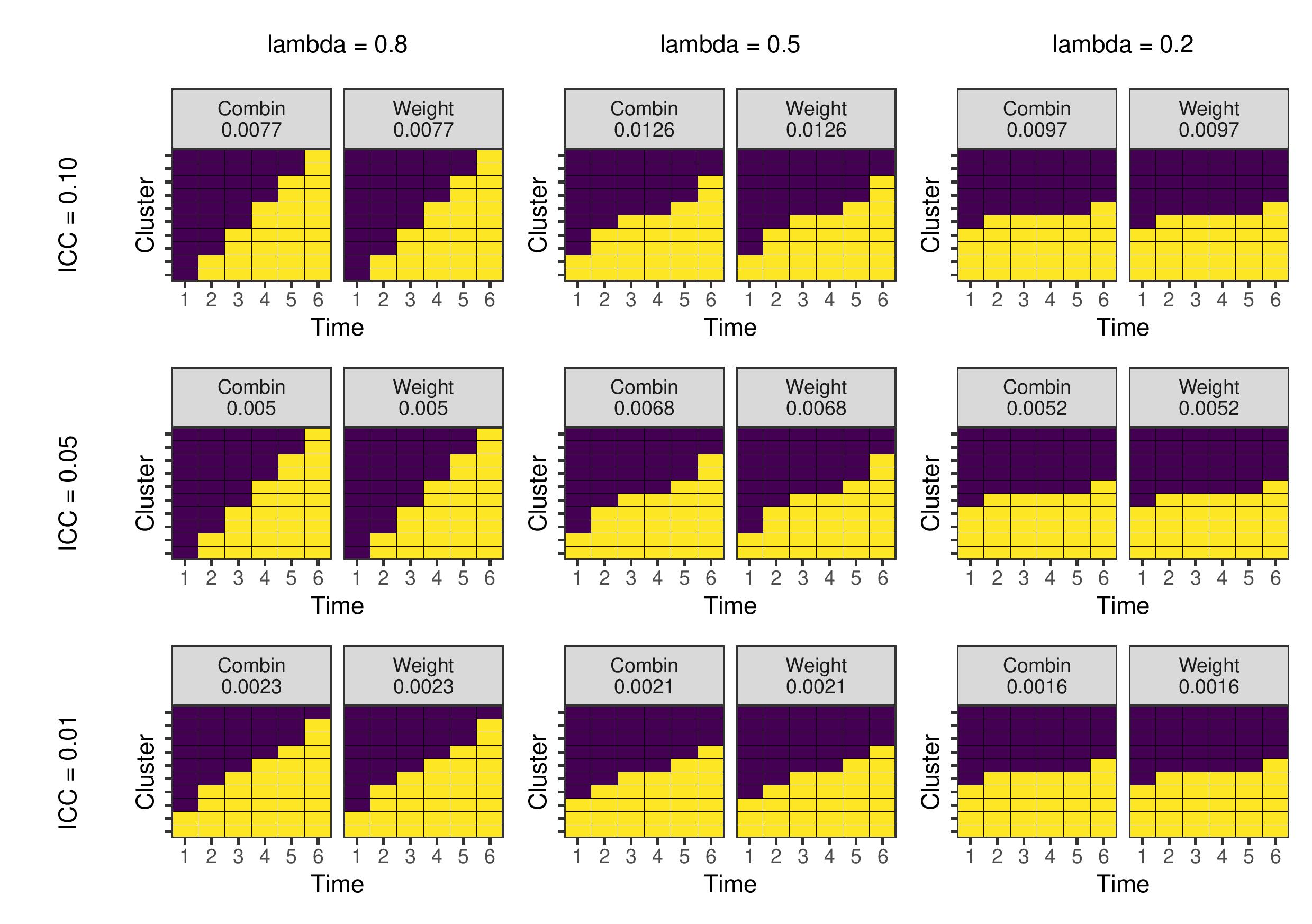}
    \caption{Optimal study designs with ten clusters and six time periods for different values of the ICC and autoregressive parameter $\lambda$ (`lambda') using a linear mixed model with AR1 covariance structure with $m=100$ individuals per cluster-period. `Combin' are results from the combinatorial local search run 100 times and selecting the best design, `G-H' are results using the method from Girling and Hemming, and `Weight' are designs produced by estimating experimental unit weights.}
    \label{fig:ex2b}
\end{figure}

\end{document}